\documentclass[conference,compsoc]{IEEEtran}
\usepackage{xr}
\usepackage{rotating}
\usepackage{framed}
\usepackage{tabularx}
\usepackage{booktabs} 
\usepackage{multirow}
\usepackage{xcolor}

\newcolumntype{L}{>{\raggedright\arraybackslash}X}

\ifCLASSOPTIONcompsoc
  \usepackage[nocompress]{cite}
\else
  \usepackage{cite}
\fi
\ifCLASSINFOpdf
\else
\fi
\usepackage{graphicx}

\begin{document}
\author{\IEEEauthorblockN{Shuning Zhang\IEEEauthorrefmark{1},
Hui Wang\IEEEauthorrefmark{2},
Rongjun Ma\IEEEauthorrefmark{3}, 
Xin Yi\IEEEauthorrefmark{1},
Kanye Ye Wang\IEEEauthorrefmark{4},
Robert Xiao\IEEEauthorrefmark{5} and
Hewu Li\IEEEauthorrefmark{1}}

\IEEEauthorblockA{\IEEEauthorrefmark{1}Tsinghua University\\ 
Email: zsn23@mails.tsinghua.edu.cn}

\IEEEauthorblockA{\IEEEauthorrefmark{2}University of Duisburg Essen\\
}

\IEEEauthorblockA{\IEEEauthorrefmark{3}Universitat Politècnica de València\\
}

\IEEEauthorblockA{\IEEEauthorrefmark{4}University of Macau\\
}

\IEEEauthorblockA{\IEEEauthorrefmark{5}University of British Columbia\\
}}


\title{Understanding Security and Privacy Perceptions of Content Creators Regarding AI Labels of AI-generated Content} 
\maketitle

\begin{abstract}
AI labels, typically implemented via underlying tracing mechanisms such as watermarks and metadata, are crucial for protecting Artificial Intelligence-Generated Content (AIGC) against security threats like disinformation and evasion. 
However, the perceived devaluation of AI-assisted work discourages creators from disclosing AI use, incentivizing efforts to bypass labeling and compromising downstream traceability.
Yet, how AIGC creators perceive the security and privacy (S\&P) implications of these labels, and how their behaviors impact technical resilience remain underexplored. To this end, we conducted semi-structured interviews with 21 AIGC creators and measured images across 6 image generation platforms against 16 self-reported manipulation settings. Our findings reveal that creators conflate binary AI labels with granular traceability, and express strong fears of de-anonymization via platform identifiers. Driven by fears of algorithmic traffic suppression and reputational risks, they defensively removed digital traces. Through empirical tests, we show that targeted modifications like coarse quantization significantly degrade detection. AI detection capabilities are also inconsistent across platforms, and suffer from false positives even for human-authored images. Based on these insights, we advocate for workflow-resilient implicit AI labels that align technical guarantees with creators' incentives.

\end{abstract}


%
\IEEEpeerreviewmaketitle

\section{Introduction}

Generative Artificial Intelligence (AI) tools have become integral to modern digital workflows, where creators increasingly adopt them for content generation and refinement. From a survey of Adobe, 86\% of creators are actively using AI in their workflow\footnote{https://news.adobe.com/news/2025/10/adobe-max-2025-creators-survey}. To avoid model evasion and trace disinformation, platforms, such as OpenAI\footnote{https://openai.com/index/advancing-content-provenance/} and Google\footnote{https://deepmind.google/models/synthid/}, implement \textbf{AI labels}, which are technically supported by underlying tracking technicalities like watermarks and metadata embedded within AI-Generated Content (AIGC). 
This paper scopes AI labels as any markers affixed to AIGC for purposes such as safeguarding copyright, proving authorship, or distinguishing it from human-authored content, including both platform-added and user-added markers.
These labels serve to enhance data provenance~\cite{abdelnabi2021adversarial}, facilitate copyright protection~\cite{guo2025audio}, and mitigate disinformation spreads~\cite{qu2025provably}.

However, these labels introduce a dilemma for \textbf{AIGC content creators}, \textit{defined as individuals or entities leveraging generative AI tools to co-create, augment, or refine digital content}. On one hand, the presence of these labels causes audience to devalue the creators' work, scrutinizing it as low-effort or low-quality~\cite{rae2024effects,jung2025ai}, incentivizing creators to circumvent or obfuscate these AI labels~\cite{zhang2024confrontation,messer2024co}. On the other hand, the absence of AI labels leaves users unable to discern synthetic content, and leading to disinformation dissemination~\cite{kasra2018seeing,wolf2026seeing}.
Yet, little is known about whether AIGC creators are aware of the potential consequences of manipulating \textbf{AI labels}, how they practically evade or alter such label mechanisms, and how they perceive the security and privacy (S\&P) implications of these labels. Moreover, the robustness of existing AI label mechanisms under real-world creator manipulations remains underexplored.
To this end, we investigate the following research questions (RQs):

$\bullet$ \textbf{RQ1:} How do AIGC creators perceive the purpose of AI labels and their associated  S\&P risks?
%

$\bullet$ \textbf{RQ2:} What practices do AIGC creators use to remove, retain, or add AI labels when sharing AI-generated content?

$\bullet$ \textbf{RQ3:} How do these practices impact the efficacy of AI labels, and what are the S\&P implications?

We conducted semi-structured interviews with 21 experienced AIGC creators, and evaluated AI label detection capabilities across 6,000 images from 6 models against 16 self-reported manipulation flows.

Towards RQ1, we identified several misunderstandings among AIGC creators. While they correctly recognize S\&P functionalities of AI labels, such as mitigating deepfakes and facilitating digital forensics, they conflate binary AI labels with granular traceability traces. Furthermore, they overestimate the privacy risks associated with generic platform AI labels, fearing de-anonymization via IP or identifier leakage, while lacking clarity on how opaque algorithmic detection triggers false authenticity penalties and shadowbanning.

Towards RQ2, we categorized creator behaviors regarding AI labels into three practices: removal, retention, and addition. AIGC creators \textit{removed} visible AI labels to preserve professional credibility and circumvent platform moderations. They \textit{retained} labels to ensure regulatory adherence and mitigate reputational backlash. They proactively \textit{added} AI labels, such as zero-width characters and personal signatures, to defend against automated data scraping and intellectual property theft.

Towards RQ3, we found that commercial AI label detectors generally maintain stable score calibrations across standard modifications, with significant degradation occurring only under coarse color quantization. Conversely, metadata is highly fragile under manipulation, as evidenced by C2PA and IPTC detection rates dropping from to near 0\% with center-cropping or recompression, compared with original images. Despite these metadata vulnerabilities, invisible watermarking techniques, such as Stable Signature~\cite{metaai2023stablesignature} and Jonathan's detectors~\cite{clark2025imagewatermark} signal show high survivability under crop and JPEG compression manipulations, though resizing reduces detection capabilities.

Collectively, the contributions of this paper are threefold:

$\bullet$ We provide the first analysis of AIGC creators' mental models regarding AI labels' S\&P benefits and risks, revealing their misunderstandings. 

$\bullet$ We categorize their label manipulations practices into three categories and revealed underlying behavioral incentives.

$\bullet$ We empirically validate the effects of their workflows on AI label detection, revealing vulnerabilities and cross-platform inconsistencies.

\section{Background and Related Work}


In this section, we first introduce creators' workflow to contextualize the practices around AI labels. We then synthesize AI labels, content provenance and watermark techniques. We finally present works around users' perspectives on these techniques.

\subsection{Creators' Workflow}

\begin{figure}[h]
    \centering
    \includegraphics[width=\linewidth]{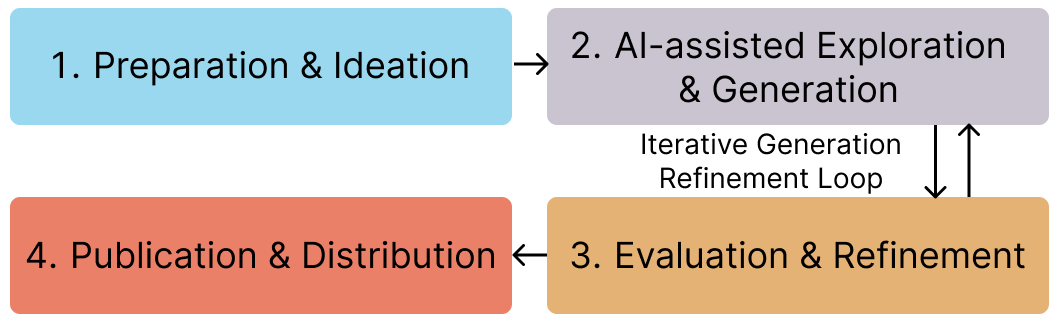}
    \caption{Creators' workflow with generative AI.}
    \label{fig:creator-workflow}
\end{figure}

%
Prior literature suggests that generative AI assists creators across different aspects of the creative process~\cite{zhang2025agency,liu2022design,chen2026generative,wan2024felt}. Accordingly, we identify four stages of AI-assisted workflows, as illustrated in Figure~\ref{fig:creator-workflow}.
First, creators engage in ideation and material preparation, translating their abstract goals into concrete prompts. Second, they use AI tools to explore the candidate output space, such as sample images. Third, creators evaluate and manually refine these outputs. When results misalign with their intent, creators edit the artifact, adjust system parameters, or reformulate prompts, creating an iterative loop between generation and refinement steps. Finally, creators review, publish, and distribute the finalized work across target channels. While some creators modify content post-publication, these adjustments remain within the scope of the workflow we described. Collectively, these stages describe the creation and manipulation flows central to our research, where providers embed AI labels during initial generation, and downstream edits by creators manipulate these AI labels.


\subsection{AI Labels, Content Provenance and Watermarking}

AI labels serve as the primary user-facing layer for content attribution and transparency of AIGC, relying on underlying technical indicators like watermarking and metadata. These technical traces act as critical mechanisms for copyright protection and algorithmic traceability across online information systems~\cite{howison2011validity, young2026media, nemecek2026authenticated}. They are broadly categorized into explicit (user-perceptible) watermarks~\cite{bistron2026deep, hwang2023brief}, implicit (imperceptible) watermarks~\cite{darwish2024blockchain, meng2026advances}, and metadata embedded directly within image fields~\cite{bushey2025cryptographic, nin2013digital}. Functionally, these components provide vital S\&P capabilities, including protecting creators from copyright infringement~\cite{guo2025audio, zhao2021structural}, safeguarding proprietary data against model extraction~\cite{lv2024mea, pegoraro2024deepeclipse}, enabling forensics and traceability~\cite{pegoraro2024deepeclipse}, and mitigating the spread of disinformation~\cite{qu2025provably, zong2025audiomarknet}.



S\&P literature found that metadata enhances information security and data ownership verification, yet uncontrolled exposure introduces multi-faceted privacy threats. On the utility side, metadata effectively protects image integrity against cropping, screenshotting, and social media compression~\cite{bin2017study}, legalizes business compliance traceability~\cite{munier2013legal}, unifies cross-domain research schemas~\cite{ulrich2022understanding}, elevates public trust in data visualizations~\cite{burns2021making} and acts as a robust authorship safeguard when integrated with physically unclonable functions (PUFs)~\cite{zhang2025hard}. Conversely, public metadata exposes severe risks, where geographic fields can pinpoint residential addresses or infer sensitive visits~\cite{drakonakis2019please, kumar2016location, baron2020you}, telephone metadata allows high user re-identifiability~\cite{mayer2016evaluating}, and Twitter metadata can uniquely fingerprint identities via machine learning~\cite{perez2018you}, though architectural dependencies such as federated learning do not necessarily exacerbate this leakage~\cite{zhan2024will}. \textbf{Despite extensive analyses of metadata utility and risks, existing papers overlook creator-side behaviour within generative AI contexts.}

Current content provenance frameworks primarily rely on either actively embedding traces, or passively verifying the files. Active embedding methods integrate provenance signals into the media through user-facing visible watermarks~\cite{chen2026generative}, or invisible techniques such as steganography~\cite{liang2026watermarking}, frequency-domain watermarks~\cite{zhao2024invisible}, and cryptographically signed metadata manifests~\cite{c2pa_advancing}. Conversely, passive verification uses server-side infrastructure, including perceptual hashing and centralized content analysis, to match and trace assets without modifying the underlying files~\cite{steinebach2023analysis,mohit2026provenance}. Standard implementations of these methods include Google's SynthID~\cite{dathathri2024scalable,gowal2025synthid} and the C2PA specification~\cite{c2pa_advancing}. Technical evaluations via benchmarks like WAVES show that while certain signals survive routine pixel-level transformations~\cite{an2024waves,zhao2024invisible,jovanovic2026watermarking}, invisible watermarks frequently fail under intentional removal attacks~\cite{kassis2025unmarker,shamshad2025first,zhao2023provable}, and mainstream social platforms routinely strip embedded metadata during upload~\cite{moruzzi2025content,dhawka2025data}. \textbf{However, these evaluation focus on algorithmic robustness, leaving a gap in how the perceptions and behaviors of AIGC content creators impact real-world AI labels.}

Usability research in S\&P domain often shows that users misinterpret S\&P mechanisms, revealing a disconnect between technical guarantees and user expectations. Prior literature on HTTPS interfaces, SSL warnings, and secure messaging protocols indicates that users routinely over-rely on superficial visual cues, and misalign their behavioral expectations with the mathematical realities of underlying communication protocols~\cite{sunshine2009crying, krombholz2019if, abu2018exploring, unger2015sok}. This cognitive misalignment extends to data lifecycle management, where individuals underestimate the persistence of residual metadata and information traces following deletion commands~\cite{abu2020designing, shahriari2026systematic}. \textbf{Our work extends this line of inquiry into the rapidly evolving AIGC landscape, exploring the disconnect between AI creators' mental models and the technical realities of AI labels.}

\subsection{AIGC Creator and Viewers' Perceptions for Provenance Techniques}

Prior work investigated user perceptions of provenance techniques from creator, viewer, and societal perspectives. 

From creator perspectives, research investigated user perceptions and behavioral manipulations regarding metadata~\cite{tayeb2018toward,henne2014awareness}. Users often show varying degrees of awareness and distinct behavioral patterns when managing hidden digital traces. While some leverage tools to proactively strip sensitive metadata for privacy defense~\cite{tayeb2018toward,henne2014awareness}, others treat historical metadata as cultural artifacts for digital commemoration~\cite{jorgensen2023designing}. Tensions in metadata between technical utility and personal privacy drive researchers to propose collaborative profiles~\cite{coleti2020tr}, localized alert services~\cite{henne2013snapme}, and privacy-preserving metadata obfuscation frameworks~\cite{chen2024scalable}, alongside design principles around metadata exposure that is tailored to user attributes~\cite{beard2018digital,furini2015location}. However, they do not explore how AIGC creators understand and manipulate provenance methods like AI watermarks. 

From viewer perspectives, literature examines how viewers perceive and process AI labels to evaluate content authenticity. The presentation and granular details of AI labels significantly shape audience cognitive processing and verification behaviors. For instance, detailed authorship labels can enhance trust and perceived human contribution in co-created content~\cite{xiao2025authorship}, while standardized process-based transparency labels help viewers evaluate content authenticity and mitigate image-based misinformation~\cite{burrus2024unmasking,holtervennhoff2026s}. To investigate these effects, researchers developed design spaces spanning sentiment, iconography, positioning, and detail levels to evaluate how warning prototypes mitigate misinformation risks on social platforms~\cite{gamageLabeling}. Unlike these viewer-centric papers, we focus on how creator perceive and interact with AI labels.

From societal perspectives, users generally prefer creator self-disclosure over automated detection and distinguish between ``AI-modified'' vs. ``AI-generated'' based on perceived human involvement~\cite{jung2025ai}. Disclosing AI assistance were found also to reduce audience satisfaction and harm creator perception~\cite{rae2024effects}. These findings motivate examining how creators manage AI label functionalities and the potential adverse consequences.

\section{Methodology}

This research was approved by our university's Institutional Review Board (IRB). We first conducted semi-structured interviews with AIGC creators, followed by a technical evaluation assessing image robustness against manipulations self-reported by participants. 

\subsection{Interviews on AIGC Creators}

\textbf{Interview design and procedure.} We conducted semi-structured interviews around creators' S\&P concerns and practices within the AIGC ecosystem. The interview protocol was structured to examine the socio-technical tension between platform-mandated AI labels and creators' behaviors across four aspects: (1) perceived S\&P threats and implications, (2) practices regarding whether, how and why content creators retain, manipulate or add AI labels, (3) challenges and expectations related to AI labels, content provenance, and associated S\&P implications. We conducted interviews via video-conferencing software (i.e., Zoom, Tencent Meeting) using Institutional accounts in Chinese, English, or German, depending on participants' preference. All sessions were audio-recorded with the explicit informed consent of participants, and transcribed verbatim. Transcripts were verified by authors to ensure accuracy.

\textbf{Recruitment.} Participants were recruited via social media (i.e., WeChat, RedBook, Whatsapp) and institutional channels (i.e., university student groups) following screening guidelines by Panicker et al.~\cite{PanickerFraudulence}. Eligibility criteria required participants to (i) post at least twice per week, indicating regular AIGC use, and (ii) maintain an active public account with at least 100 followers. We recruited 21 active AIGC creators between November and December 2025. 
The participant sample included creators based in China ($n=13$), Germany ($n=7$), and Italy ($n=1$), 
covering image, video, and multimedia social media creators who regularly distribute AI-generated or AI-assisted content across major content-sharing platforms (e.g., TikTok, Youtube, Instagram, Redbook). These regions represent two highly active yet divergent ecosystems regarding AIGC usage and AI label regulatory frameworks. Their educational backgrounds range from high school to Ph.D. levels, and detailed demographics are shown in Table~\ref{tab:demographic}. 

\textbf{Data analysis.} Three researchers independently analyzed an initial subset of transcripts (3 scripts), and iteratively refined the codes through cross-check and consensus discussion. 
As our coding process was inductive and interpretive, we did not calculate inter-rater reliability. Instead, we used consensus discussions to resolve differences, and ensure coherence in the application of codes~\cite{mcdonald2019reliability}.

\subsection{Measurement Study}

To evaluate the real-world resilience of AI labels, we examined 16 technical manipulations from the high-frequency workflows reported by creators (Sec.~\ref{sec:practice}). While some text-based methods (e.g., zero-width characters) emerged in interviews, we focus on image domains, where AI labels are most frequently deployed. By grounding evaluations in empirically observed user behaviors rather than synthetic attacks, we benchmark the viability of current AI labels, especially under manipulations.


\textbf{Model selection and dataset generation.} 
We selected six representative text-to-image models with diverse brands and regions: Midjourney, Jimeng, Wenxin Yiyan, Qwen, ChatGPT and Gemini. We opted not for video generation models as users reported negligible engagement with video manipulation workflows.

For each model we used different prompts from PromptHero and Lexica prompt store to generate 500 images each. These prompt stores are commonly used in prior S\&P works~\cite{deng2026prompt,shen2024prompt}, and also recommended by the public\footnote{https://influencermarketinghub.com/ai-prompt-marketplaces/}. As reported, PromptHero has over 1.1M monthly visits in Apr 2026, and feature millions of prompts\footnote{https://www.semrush.com/website/prompthero.com/overview/}. Lexica has 5M+ prompts~\cite{shen2024prompt}. We selected prompts to diversify the topics, covering multiple domains such as animals, anime, architecture, fantasy, food, landscape, portrait, and science fiction. These processes totalled 6,000 images, with 1,000 images for each model (e.g., ChatGPT).
Besides, to quantify false-positive risks, we evaluated on additional human-authored baseline datasets compiled from NASA and Wikimedia repositories, totalling 1,000 images. 

\textbf{Detector selection.} We selected 3 commercial AIGC detectors ( aiornot\footnote{https://www.aiornot.com/}, illuminarty\footnote{https://illuminarty.ai/de/}, and hivemoderation\footnote{https://hivemoderation.com/}), two representative watermark or metadata detection tools (Stable Signature\footnote{https://ai.meta.com/blog/stable-signature-watermarking-generative-ai/} and the tool by Jonathan Clark, denoted Jonathan\footnote{https://jonathanclark.com/posts/image-watermark-detection.html}), and two manifest-parsing standards (C2PA and SynthID). Crucially, as commercial AI detectors operate primarily as black-box classifiers rather than low-level signal extractors, our evaluation focuses on downstream detectability to reflect real-world platform deployment.



\textbf{Manipulation method selection.} We selected the following manipulations: JPEG compression, quantization, cropping, and screenshotting, following prior practices~\cite{lukas2023ptw}. Details are described in Appendix~\ref{app:para_details}. To isolate the effect of each manipulation, we did not combine them, since any compromise of watermark or metadata in an earlier stage would propagate through all subsequent stages.

$\bullet$ JPEG compression: images were compressed across varying quality factors ($q \in [80, 100]$), following prior practices~\cite{lukas2023ptw} and participants' self-reported practices.

$\bullet$ Color quantization: each image's RGB channel is normalized to [0,1], then quantized as $\hat{v} = q\cdot \lfloor \frac{v}{q} \rfloor$, which reduces the number of color levels. We use four step sizes $q \in \{0.01, 0.02, 0.05, 0.10\}$, inspired by prior benchmark~\cite{lukas2023ptw}. 


$\bullet$ Cropping and resizing: Images were center-cropped at different ratios ($\rho \in [0.9, 1.0]$), and subsequently upscaled back to their original sizes. 

$\bullet$ Screenshot: We screenshot images at their original resolution, under macOS and Windows environments. The results reported is the average of multiple screenshots.

\textbf{Detecting aspects.} For each image, we evaluated digital tracing methods across their target fields: 

$\bullet$ AI-generated: as most commercial platforms only return AI-generated or not, we evaluate whether an image is AI-generated using commercial detectors.

$\bullet$ Metadata: we target signed provenance manifests (i.e., C2PA standard), extracting the \texttt{claimSignature}, \texttt{generator} tool name, and \texttt{instanceID}. We also extract generative Exchangeable Image File (EXIF) and text chunk parameters, including model versions, user IDs, and prompt seeds. 

$\bullet$ Watermark: although most tools and commercial detectors do not feature watermark detection, we embed specific watermarks, such as Meta's Stable Signature~\footnote{https://facebookresearch.github.io/meta-seal/}, and detect this specific watermark using Meta's tool.

\textbf{Analysis.} Statistical significance across models and platform variations was calculated via Pearson $\chi^2$ tests and two-proportion $z$-tests. For continuous detection score distributions under baseline and attacked conditions, we performed paired exact McNemar tests, two-sided Mann-Whitney $U$ tests, and two-way analysis of variance (ANOVA) with post-hoc contrasts. Descriptive results are reported with 95\% Wilson confidence intervals.

\section{RQ1: Perceived S\&P Benefits and Risks of AI Labels}



This section explores how creators conceptualize AI labels. We found while creators recognize the perceived S\&P benefits of these labels, they also thought AI labels have S\&P risks. This tension influences their subsequent behaviors and practices. Although our study focused on AI labels, participants often discussed AI labels alongside other creator-applied labels, such as logos and copyright notices. We include these examples as they show how creators understand AI labels within broader practices of signaling authorship, ownership, and authenticity.

\subsection{Perceived S\&P Benefits}

Participants identified seven S\&P benefits associated with AI labels.

\subsubsection{Deepfake, fraud and deception prevention (13/21 mentions)} 
Participants thought that AI labels act as mechanisms against AI-driven deepfake exploits, such as scams, identity spoofing, and malicious frauds. They highlighted that in generative AI era, distinguishing real content from deepfake ones is challenging (P15). This risk is particularly concerning regarding susceptible groups.
As P11 noted, \textit{``I think it might mislead older demographics into believing it is real ... If there are not AI labels, aren't there many people using AI to face-swap for portraits now?''} 
Beyond these groups, the vulnerability also extends to social media platforms.
P15 said, \textit{``Since these images no longer carry AI labels, it has become highly challenging for people on platforms like Instagram to discern whether there is a real individual or a fabricated persona behind an account.''} 
Besides visual deepfakes, participants worried about audio-based deepfake, especially voice cloning for targeted frauds. As P08 shared, \textit{``[A] serious concern is voice [cloning] ... particularly in phone call scenarios, like fake calls where [scammers] use someone's voice to deceive their family members ... it is quite a terrible scenario.''} 

Beyond deepfakes, participants highlighted deceptions. P16 observed that short-video platforms and digital art platforms frequently use highly realistic AI-generated content to mislead the audience, such as presenting fake news or AI-generated work that mimic original artists to deceive viewers.
Participants further noted that unlabeled synthetic media show deceptive tutorial content that depicts physically impossible processes (P18) or graphic violence (P19). For example, P11 remarked, \textit{``Without adding AI labels, some images could be misleading... making people believe the content is real, but the steps are impossible to follow, or such phenomena could never occur in reality.''} 
Towards these risks, participants viewed watermarks as important tools for warning users. P13 echoed, \textit{``[watermark] helps others identify the provenance of the content.''} 

Finally, participants argued that once specific sensitive images enter the public domain, unrestricted access makes absolute prevention of misuse nearly impossible. As P15 noted, \textit{``It is almost impossible to completely prevent misuse because nearly anyone can access [them]. I think personalized watermarks might help to some extent, because it can show the image belongs to a specific person.''}

\subsubsection{Authenticity proof (7/21 mentions)} 
Participants view AI watermarks as signals to help audience verify content authenticity and differentiate synthetic content from real-world records. 
They believed that watermarking is a basic verification mechanism on public social media to counter potential fabrication (P13, P14).  
For instance, P14 explained, \textit{``It can serve as a verification of the content source and a tool for risk mitigation.''} 
Participants also highlighted invisible watermarks' usage in judging the authenticity for derivative content.
P05 acknowledged that, \textit{``[platform developers] embed something into the underlying code ... it is invisible, but during the final adjudication, a decoding process reveals that code, proving you used it.''} 

Furthermore, participants interpreted AI labels as essential to bridge audience media literacy about authenticity (P21). They argued that watermarks preserve the threshold between reality and synthetic fabrication (P03, P16, P20). Expressing this perspective, P16 noted that \textit{``watermarks clarify whether media is AI-created content, rather than something photographed or drawn in real life.''} Similarly, P03 asserted, \textit{``if an image is presented as a photograph or documentary record despite being AI-generated, audiences should be informed, because generated content should not blur the boundary between reality and creation.''} Similarly, P20 argued that \textit{``AI-generated labels help people distinguish which things are real and which are fake.''}

However, participants thought current AI labels could not serve authenticity purposes, as they primarily indicate whether or not the image is generated by AI tools. P03 said, \textit{``watermark merely serve as a binary indicator of AI involvement ... fail to quantify the specific degree of human versus AI contribution ... current watermark infrastructures cannot establish granular provenance or delineate true authorship.''} 

\subsubsection{Prevent proprietary data from unauthorized use (11/21 mentions)} 
Participants indicated that AI labels may prevent proprietary data, such as conceptual design ideas, from potential leakage, particularly given creators' lack of transparency regarding downstream data usage (P03). This protective function also extends to biometric information. As P11 said, \textit{``without watermarks, people frequently use AI face-swapping to generate portrait photography ... where [the] original images [were] never authorized.''} These vulnerabilities are especially acute in the e-commerce sector, where unwatermarked images of human models are exploited to generate synthetic promotional content, deceiving users and compromising the human models' privacy.  

These practices are especially evident when images themselves contain users' personal data. As P11 shared, \textit{``If I used my own facial photos ... or if it contains personal information such as my school affiliation ... I would add watermarks to prevent others from misappropriating it.''} 
However, some participants thought the misuse of these famous person's proprietary data is complicated, especially for adding watermarks. \textit{``Since the faces belong to them, yet the images were generated by us, I am uncertain about how copyright should be attributed ... [and] whether this constitutes a misappropriation.''} (P18) 

\subsubsection{Tamper detection (2/21 mentions)} 
Participants worried about the potential that audience may manipulate their work, especially since there work are publicly distributed. They thought those can be mitigated by adding watermarks with their IDs, or tiny markers. As P04 mentioned, \textit{``If our videos were downloaded and maliciously modified ... I would be quite worried that the content I generated ... would be misappropriated by others.''}

\subsubsection{Forensics (5/21 mentions)} 
Participants emphasized the forensic usage of AI labels, noting their role as critical evidence during infringement disputes (P06), and as a defense against unwarranted reputational damage. P12 shared having experienced forensics risks, and shared that \textit{``it turned out to be ... a blessing ... because [the watermark] completely contradicted ... rumors.''} 
Participants also mentioned using AI labels to trace back to the creators creating harmful content. For example, P08 shared that, \textit{``If someone plagiarizes another's voice to create [harmful] content, we can trace exactly who generated this fake information.''}

Conversely, some participants argued that current AI labels can not serve forensics purposes, as they only identify the generative model rather than the content creator. As P2 argued, \textit{``Platform-level watermarks [...] do not link with individual user identities. Because they are uniformly embedded [...] they remain user-agnostic.''} 

Some participants with higher technical literacy also highlighted the importance of forensics for harmful content, especially by regulatory bodies. P12 commented, \textit{``harmful content [...] demands traceability [so that] cyber police can perform provenance tracking to hold responsible entities accountable.''} 

\subsubsection{Data stealing and extraction prevention (7/21 mentions)}
Participants thought that AI labels can deter unauthorized data extraction and content theft during external submissions, competitions, or in e-commerce scenario, especially when distributing these images to untrusted entities (P06, P09).
P06 explained that, \textit{``[Our] exports for exhibitions or competitions [...] all contain invisible watermarks. Because of this, it is rare for people to pirate them.''} 
Additionally, participants described AI label techniques to disrupt automated scraping (P12). For example, P12 described \textit{``embedding zero-width characters in the text ... so that when scrapers copy the text, their token count exceeds the limit, making AI article spinning impossible''} 

At the platform level, participants observed that service providers leverage AI labels to safeguard proprietary generative resources. As P10 said, \textit{``Within tools like Suno, an audible identifier dynamically interrupts the track ... to declare the generating platform's identity, and the asset remains bound to these constraints unless the user subscribes to a premium membership.''} 

Furthermore, participants highlighted that AI labels increase economic cost and labor required for these labels' removal and subsequent reuse. P11 expressed this friction, stating that \textit{``with a watermark, removing it would be troublesome, so it becomes harder to steal.''} They also connected the increased removal difficulty to a reduction in economic viability of stolen content. As P09 observed, \textit{``some watermarks have an anti-attack role because they are hard to remove and leave visible traces on reused images.''}

However, participants noted that AI labels functions primarily to raise the cost of unauthorized copying rather than an absolute security solution. P02 noted that although personal markings successfully discourage plagiarism by increasing AI labels' removal times, this protection remains unreliable.

\subsubsection{Copyright protection (4/21 mentions)} 
A commonly mentioned, but less privacy-correlated aspect is copyright protection, where participants thought AI labels could help prevent copyright infringement. As P05 thought, \textit{``when actual infringement occurs, watermark serve as tags or symbols to verify ... who the original creator is, which can to some extent help reduce infringement.''} 

However, participants highlighted that generative AI introduces more ambiguities regarding ownership attribution and platform compliance. Specifically, AIGC platforms often embed AI labels that creators are obligated to retain under the Terms of Service. Participants argued that this restriction compromises their agency to claim copyright rights over further modifications. As P06 said, \textit{``Some tools prohibit watermark removal ... [and] even if I heavily modify the content, it remains unclear whether the ownership belongs to me, or the platform.''} 
Some participants mentioned unauthorized voice cloning and misappropriation of vocal identities, causing copyright infringements to celebrities. P08 mentioned, \textit{``I watched a video on a video-sharing platform where someone cloned a well-known pop singer's voice to perform another artist's songs ... the video was clearly unauthorized by the original singer.''} 

\subsection{S\&P Risks}

Despite the recognized benefits, participants also recognized several S\&P concerns regarding AI labels.

\subsubsection{Data leakage (10/21 mentions)}
Participants fear that visible AI labels containing usernames or IDs can be exploited by unknown entities to link these data into their identities, causing data exposure. 
P09 said, \textit{``When watermarks contain real names or usernames ... others might find their [other] contact information. I feel this constitutes privacy exposure.''} 

Participants differentiated between low-risk public data and high-risk personal identifiers, where their perception is based on the specific category of information embedded in the AI label. For example, P19 welcomed a platform watermark to promote their content but restricted other identifiers, stating, \textit{``do not write my IP address, my real name, or my email.''} P21 also highlighted the importance of network location data, where \textit{``an IP address is a very important issue. On the internet, an IP address is like an ID number.''}

Participants therefore thought that AI label has a trade-off between functionalities, such as forensics, and privacy exposure. As commented by P13, \textit{``watermarks themselves may inadvertently encapsulate sensitive user information ... [there] needs to be a balance ... without precipitating over-exposure.''} 

\subsubsection{De-anonymization (6/21 mentions)} 
Participants' fear of data leakage further extend to a feeling of de-anonymization, as all their work have similar AI labels. They thought that the AI labels are similar to their social media ID, which could potentially be traced. P04 said, \textit{``[others] can leverage my unique watermark to trace back to my account information ... the risk arises as these digital accounts are linked to [their] personal information.''} 

\subsubsection{False authenticity (8/21 mentions)} 
Participants found platforms use opaque automated detection to mandate AI labels, creating risks when hidden or misclassified content triggers severe algorithmic penalties or shadowbanning. Specifically, they expressed apprehension that, if they fail to watermark AIGC, it could result in sudden suppression once detected by platforms. P02 remarked, \textit{``If you don't tag [watermark], your work might be quite good ... But once they discover it, your post or video is [banned].''} P21 similarly observed that when content is labeled as AI-generated, creators faced anxiety regarding the sudden reduction of platform traffic. 

Participants emphasized that these vulnerabilities are compounded by the unreliability and biases of AI detection tools. They noted that automated tools fail to provide accurate assessments of content origin, frequently misclassifying human contributions. P13 shared that after they manually polished the text, the detected AI rate paradoxically increased. Furthermore, P21 argued that AI detectors often yield highly contradictory results, meaning that there is not yet a clear, unified standard for judging whether content is human-made or AI-generated.

They further argued that mislabeling compromises trust, whether images falsely claim human authorship or is erroneously flagged as synthetic. P20 stated that if AIGC is mislabeled for human-created work, audience raises doubts about authenticity and misunderstand how the content was produced. Similarly, P21 echoed that audience may dislike these work, or feel deceived.

\subsubsection{No risks (10/21 mentions)} 
Participants thought there are negligible privacy risks, as AI labels do not contain proprietary data. 
As P14 said, \textit{``If the watermark contains my account information -- like my Xiaohongshu ID -- it doesn't really touch upon deep personal information ... I don't think it poses any risk of identity exposure.''} 
Participants further observed that platform labels primarily serve as disclaimers than tracking mechanisms, conveying no user information. For instance, P01 noted, \textit{``platform AI watermarks usually only say that content was AI-generated and do not contain personal information.''} Consequently, participants believed that this prevents downstream viewers from inferring any private information, though tracing origins also becomes difficult. As P02 explained, \textit{``from a platform watermark alone, viewers have no way to judge what connection it has to your personal identity ... [or] whether it is plagiarism.''} 

\section{RQ2: Practices Around AI Labels}\label{sec:practice}

Creators' practices towards AI labels contain three aspects: remove, retain or add labels. To examine these practices and uncover behavioral patterns, our analysis examines \textit{what} types of labels are targeted, \textit{in what context} these practices occur, \textit{when} creators have these practices, \textit{how} they are implemented, and \textit{why} creators perform these practices.

\subsection{Remove AI Labels}

\subsubsection{What} Participants primarily remove visible AI labels, including corner labels applied by generative platforms (17/21 mentions), logos embedded by specific AI tools (17/21 mentions), and disclosure texts (8/21 mentions). This deletion may degrade the AI label's effectiveness to assess authenticity and track data stealing, as it remove the visual cues. 

Only 2/21 participants articulate file-level metadata stripping. When implicit or invisible marks appear in discourse, commercial pipelines often treat them as irrelevant because end audiences will not inspect bytes (P13). P07 similarly described replacing competitor logos with in-house marks without discussing hidden credentials. 

Creators mitigate detection by blending outputs from multiple AI platforms into a single project or by purchasing premium memberships to export logo-free content (P01, P02). For instance, editors routinely strip conflicting platform logos before compositing multi-source clips (P08), while premium tiers explicitly monetize watermark removal, signaling to creators that vendor marks are merely commercial branding rather than S\&P infrastructure (P02). 
P10 detailed this practices by stating, \textit{``We first remove the platform marks, and then add a unified label ... to tell everyone that the film was made with AI assistance.''}
P03 distinguished personal publishing from client delivery, \textit{``If I publish on my own channels I do not mind the watermark, but if I show it to a client I will definitely remove it ...''} 

\subsubsection{In what context} AI label removal frequently occurs in commercial (9/21 mentions), academic (3/21 mentions), or competitive contexts (3/21 mentions) where third-party branding introduces privacy risks or threats participants' professional credibility. In these environments, the presence of an AI label could expose the creator's workflow, causing biased opinions on creators, or trigger algorithmic penalties. P16 highlighted, \textit{``If it has the words AI-generated, readers may feel that it is not very convincing.''}

Within commercial work, P03 and P13 thought client- or employer-facing deliverables must not display vendor AI branding. P13 described a hard requirement not to show generation tool names on course materials sold to schools, even while acknowledging AI assistance internally. P01 reported that commercial short dramas cannot ship with platform watermarks and require an internal review step before client handoff. Within competitive work, P17 faced rules forbidding platform logos on entries. P14 worried that visible AI tags make design deliverables like cutting corners when presented to employers.

Academic and coursework contexts amplify credibility concerns. P19 removed marks before showing instructors work-in-progress renders, fearing that flawless AI output would signal low effort. P09 described instructor who get especially angry when assignments arrive with tool watermarks and actively recommend removal websites. P18 removed AI tags so follower would not discredit their content, while retaining attribution when reposting others' work.

\subsubsection{When} Removal typically occurs during post-generation editing (10/21 mentions) or at the export stage (12/21 mentions), both before publication. By eliminating these markers early in the pipeline, creators prevent tracking mechanisms across different platforms. 
P10 described removal when editing, then re-labeling at export, so stripping is often an intermediate step, not the final published state. P07 describes routinely process competitors' images before adding company logos in CapCut. P12 distinguished private Moments posts (i.e., marks removed) from public Zhihu or Xiaohongshu uploads (i.e., marks kept for compliance). P18 used photo retouching apps to erase labels before Xiaohongshu posts but skipped processing for overseas audiences.

\subsubsection{How} Participants remove labels using third-party mobile applications (12/21 mentions), desktop editing software (3/21 mentions), or other techniques like screenshotting (11/21 mentions). 
Illustrative techniques include erase using apps from the phone (P03), CapCut build-in removal and crop (P07, P09), community plugins (P01), and trim-as-last-resort (P10). P07 added a daily-limit web service (i.e., Kaipai) because CapCut re-inserts its own AI label after erasing foreign marks. P04 struggled when dense marks required paid Meitu membership and manual touch-up. P17 noted that some tools remove marks via chats or cropping. 

\subsubsection{Why} Participants removed AI labels for three reasons: manage credibility, prevent platform deprioritization, and comply with regulations. First, participants sought to manage audience perception and protect professional credibility (10/21 mentions). They fear that visible AI labels, especially watermarks, degrade the perceived value of their work and expose them to negative judgments regarding their efforts and authenticity. As P16 explained, \textit{``if it has the words AI-generated, readers may feel that it is not very convincing.''} Similarly, this reputation management extends into educational and professional scenarios, where they worry about the devaluation of their labor. P19 stated, \textit{``I would worry that the teacher thinks my attitude is not serious ... as AI can produce it in one second.''} P14 similarly removed labels because AI stamps on portfolio pieces feel like slacking or cheating even when AI only assisted drafting. 

Second, participants remove AI labels to circumvent platform deprioritization (3/21 mentions). They feared that the platforms would suppress their work if with watermarks. Participants prioritized their work's dissemination effectiveness over transparency and security. P02 highlighted the algorithm risk, \textit{``if the platform later detected AI involvement the post or video could be directly wasted.''} P07 contrasted that, while Douyin Accounts generally permit AI-generated content without throttling if properly labeled, Xiaohongshu imposes a strict ban on AI beauty videos irrespective of disclosure. P09 observed creators whose videos were taken down for suspected undisclosed AI use, reinforcing incentives to strip visible tool marks before upload. 

Finally, participants noted that removal is often enforced by institutional compliance (6/21 mentions), where they are forced to strip platform-provided watermarks to participate in further dissemination process. P17 explained, \textit{``competition works cannot use the platform's own watermark as logo.''} P13 echoed employer mandates, \textit{``my current use case has a hard requirement not to show these watermarks.''}

\subsection{Retain AI Labels}

\subsubsection{What} Participants retain platform-added disclosure watermarks (9/21 mentions), embedded metadata (5/21 mentions), and visible AI-generation tags (10/21 mentions). These are often used to preserve the authenticity signals. Retaining these elements facilitates content tracing and disinformation judgment. P10 emphasized the necessity of retaining watermarks, \textit{``AI watermarks must exist to tell your audience that this was made by AI.''} P12 treated public-platform marks as legally mandatory, contrasting them with private sharing where removal is acceptable. P15 argued that visible AI labels will become essential as synthetic media approaches photorealism. P06 highlighted the copyright protection value of invisible labeling, distinguishing it from visible corner disclosure text.

\subsubsection{In what context} Retention mostly happens in educational sharing (3/21 mentions), regulatory environments (6/21 mentions), or when communicating with audiences, where transparent disclosure mitigates privacy and authorship disputes (7/21 mentions). In these contexts, maintaining AI labels could defend against accusations of deception or plagiarism. P19 explained, \textit{``it is already AI-generated, so there is no need to hide it.''} P02 described proactively adding unified AI labels before platforms force routing into restrictive AI-only funnels after September 2024 policy shifts. P17 encouraged active labeling or provenance after seeing comment-section disputes over uncredited AI illustrations in academic work.

Fan-fiction and IP-sensitive communities complicate retention. P21 retained platform publish-time AI tags because secondary works build on others' intellectual property and should not pretend to be hand-drawn originals. P18 retained labels on celebrity likeness edits involving real people, while removing marks on merchandise-style graphics where disclosure seemed unnecessary.

\subsubsection{When} Participants retain AI labels both before publication (16/21 mentions) or during publication (15/21 mentions). They treat it as a compliance measure, and select the timestamp when the content is most vulnerable to misinterpretation or security auditing. 
P21 described publish-time labeling on fan-fiction platforms even when local downloads lack marks. P13 detailed Douyin's optional ``AI-generated'' checkbox and downloader-side account watermarks as dual safeguards that persist unless explicitly edited away. P04 noted that dynamic floating text overlays are kept through editing because they resist automated erasure than static corners. 

\subsubsection{How} Participants execute retention by accepting default platform exports (7/21 mentions) or intentionally restoring a unified mark at the end of their production (3/21 mentions). This workflow often relies on participants' own manual efforts and voluntary compliance, rather than as an automated enforcement. 
P02 designs custom overlays that are more unified than vendor defaults. P06 watermarks client drafts before contract signing to preserve attribution during revision chains. P01 accepts platform checkboxes (i.e., Douyin, Kuaishou) when posting personally, treating opt-in labels as enough for hobby content.

\subsubsection{Why} Participants retain AI labels for four reasons: fostering transparency, ensuring regulatory compliance, minimizing operational friction, and mitigating misinformation. First, they retained AI labels to maintain transparency, countering against potential authorship disputes and ethical backlash (5/21 mentions). Participants recognize that concealing AI involvement poses higher long-term reputational risks than disclosure, especially in learning environments or public-facing contexts. They retain the AI label to manage audience expectations and prevent their content from being misperceived as human-made. As P19 emphasized, \textit{``it is already AI-generated ... there is no need to hide it.''} P18 argued that labeling protects non-fans who might otherwise believe a celebrity performed actions they never did. P15 stressed that visible marks help viewers who immediately believe sensational clips and rarely revise opinions later. 

Second, participants retained AI label as a compliance strategy for platform and regulatory governance (6/21 mentions). They restore the AI label to signal rule adherence at the moment of publication. For instance, P21 expressed, \textit{``I used it, so I used it. Even if it is limited in traffic, I do not really mind.''} P12 cited mandatory explicit labeling after September 1, 2025 rules requiring 50\% opacity AI-generated badges on images. P10 said compliant creators add disclaimers even after removing vendor marks. 

Third, participants retained AI labels because it is too laborious, especially given that they think the efforts required to remove the AI label outweighs the perceived privacy or visibility risks (4/21 mentions). P01 noted this calculus, \textit{``it is not that I do not have the technique. It just feels too troublesome and unnecessary.''} P04 tolerated marks when removal would damage composition or fail on floating overlays. P08 retained marks on Jimeng posts because the content never leaves the generator platform ecosystem.

Finally, participants desired to retain AI labels as it aids in preventing misinformation by warning viewers of synthetic origins (6/21 mentions). Participants agreed that AI labels should act as a transparent boundary between synthetic and real content. P10 warned that elderly viewers may treat uncredited AI clip as real, \textit{``you must add an AI watermark to tell your audience, otherwise some audience cannot tell it is fake.''} P11 expected platforms to require labels on highly realistic generations even though they currently removes tool marks for personal Douyin posts.


\subsection{Add AI Labels}

\subsubsection{What} Participants add visible custom identity marks (e.g., logos, personal signatures, tea identifiers, 16/21 mentions), or invisible marks, such as zero-width text characters (1/21 mentions) and invisible pixel-level modifications (4/21 mentions). 
P06 commented, \textit{``a digital watermark is an ID card for our work.''} P04 layered dynamic text that moves during playback to resist erasure, while P12 embedded zero-width characters in newsletters to break automated article spinning. P17 described competition entries that carry personal implicit fingerprints rather than platform corner logos. P14 equated designer signatures with provenance markers analogous to traditional authorship seals.

\subsubsection{In what context} Participants add custom AI labels when content holds high economic value (9/21 mentions), contains sensitive personal data (2/21 mentions), or exists in an open environment vulnerable to automated scraping and security breaches (7/21 mentions). 
Commercial branding (P05, P07), client PDF deliverables (P03), and anti-scrape video overlays (P04) are high-stakes contexts that needs labeling. P13's employer adds faint corporate logos to course videos after AI asset generation. P09 watermark second-hand photos on Xianyu after screenshot theft experiences. P15 always watermark monetized social posts, noting fans would suspect sudden removal. P16 uses visible marks on commercial AI video to deter direct reposting of scripts and characters.

\subsubsection{When} These labels are added early in the drafting process before public exposure (3/21 mentions), or at the final export stage to substitute platform watermarks (8/21 mentions). Applying these protections early facilitates privacy protection before any potential leakage occur. P06 noted the urgency of this timing, \textit{``If you do not add a watermark, the first line of defense is broken.''}
Examples include watermarking drafts shared before contract finalization (P06) or embedding floating marks within the editing timeline (P04). Conversely, others defer labeling to the final deployment stage. This involves applying overlays at publication to preempt platform enforcement, exporting disclaimers after stripping vendor marks (P10), or embedding implicit marks upon competition submission while removing visible logos (P17).

\subsubsection{How} Participants add these AI labels by embedding personal signatures (10/21 mentions), restricting identity information to pseudonyms (6/21 mentions), or using text disruptions to disable malicious extraction tools (1/21 mentions). 
P04 configures CapCut text overlays that float across frames. P11 pairs optional marks with download disabling on Douyin. P21 treats historical painter signatures as precedent for modern creator stamps on fan-art reposts. P09 uses WPS document image watermarks for academic figures where erasure is difficult. 

\subsubsection{Why} Participants added AI labels because of three reasons: protect intellectual property, reduce data theft, and based on privacy calculus. First, they added labels to enforce attribution and prevent intellectual property theft (7/21 mentions). They perceive these additions as visual and technical barriers that elevate the cost of unauthorized reuse or commercial exploitation. For instance, participants found watermarks can act as an identity card containing signatures and creation dates to protect their work (P06), and that their resistance to removal helps trace the origins of deepfakes (P08). Others highlighted technical difficulties of removing. P11 explained, \textit{``removing it would be troublesome, making the work easier to steal.''} P09 described, \textit{``watermarks have an anti-attack role as they are hard to move.''}

Second, participants added AI labels to reduce data scraping and content spinning (5/21 mentions). P12 detailed, \textit{``by embedding zero-width characters in the text, their token ... exceed the limit, making AI article spinning impossible.''}

Finally, participants added AI labels after privacy calculus about identity limitation (3/21 mentions). They balanced the security benefits of attribution against the vulnerability of personal data exposure. P19 highlighted this boundary, \textit{``rather than my real name or my real email, I hope it is my ID.''} P09 articulated the threats, \textit{``If someone uses their email as a username, others might find their QQ contact information.''}

In summary, we categorized content creators' practices into four classes, as in Table~\ref{tab:workflow_mapping_compact}, which guided subsequent measurements.

\begin{table}[!htbp]
\centering
\caption{Creators' self-reported workflows.}
\label{tab:workflow_mapping_compact}
\small
\resizebox{0.47\textwidth}{!}{%
\begin{tabular}{@{}llc@{}}
\toprule
\textbf{Category} & \textbf{Self-reported workflow} & \textbf{Frequency} \\
\midrule
Spatial & Cropping / screenshotting & 13 (61.9\%) \\
Quality & JPEG / platform compression on upload & 9 (42.9\%) \\
Color Processing & Beautifying / filter adjustment & 8 (38.1\%) \\
Format & Format conversion / resizing & 4 (19.0\%) \\
\bottomrule
\end{tabular}
}%
\end{table}

\section{RQ3: Technical Robustness}

This section presented technical results of AI-generated image detection (Sec.~\ref{sec:ai_generated_detection}) and metadata detection (Sec.~\ref{sec:metadata_detection}). 

\subsection{AI-Generated Image Detection}\label{sec:ai_generated_detection}

\textbf{Accuracy across manipulations.} Commercial AI image detectors largely retain their accuracy across standard image manipulations. With the exception of severe color quantization (\textit{quantize\_q010}), which significantly degrades detection capability ($\Delta=+0.041$, $p < .001$, see Table~\ref{tab:rq3_manipulation_posthoc}), other modifications such as center cropping, screenshotting, and standard JPEG compression yield negligible shifts in detection scores ($p \ge .282$). 



\begin{figure}[!htbp]
    \centering
    \includegraphics[width=0.47\textwidth]{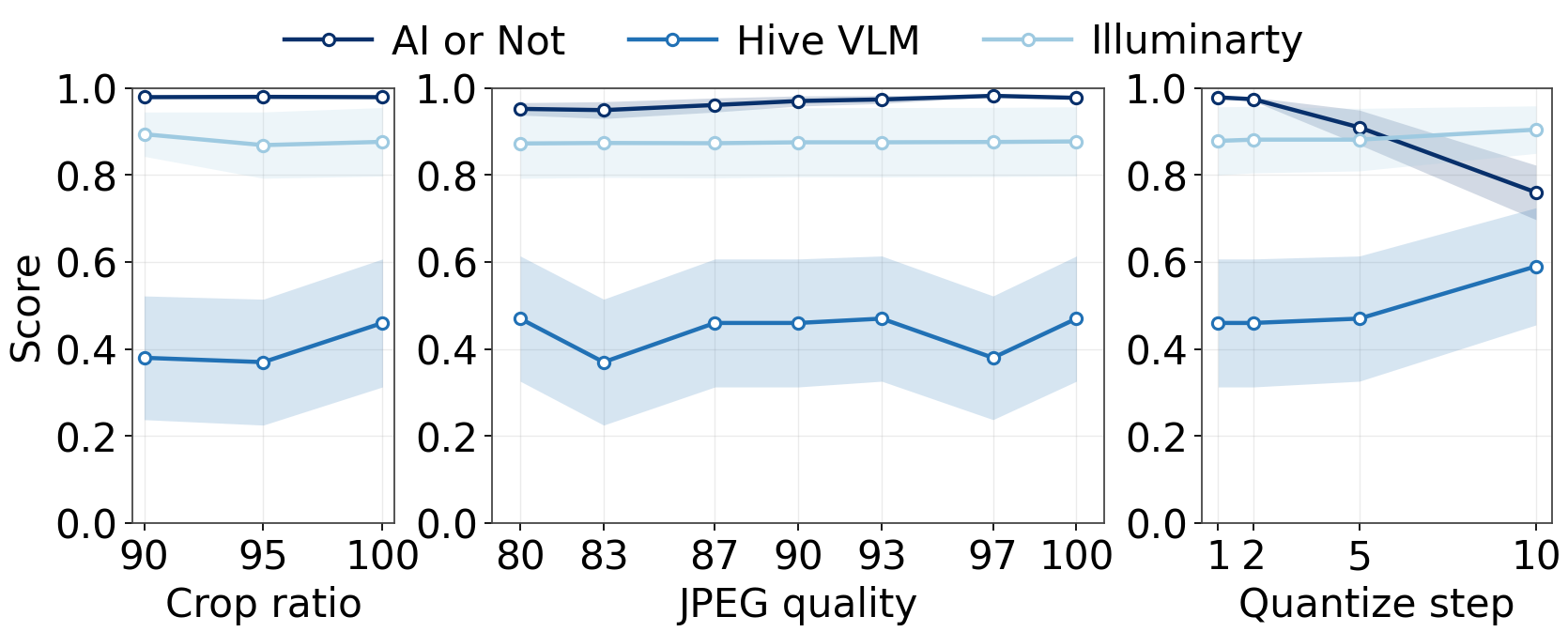}
    \caption{Scores (0=not AI-generated, 1=AI-generated) across detection platforms.}
    \label{fig:accuracy_comparison}
\end{figure}

\textbf{Performance variability across detectors.}
Detection accuracy is highly inconsistent across platforms, as in Figure~\ref{fig:accuracy_comparison}. For example, \textit{AI or Not} shows near-ceiling performance ($\mu{=}0.955$), while \textit{Hive} exhibits severe under-calibration ($\mu{=}0.448$). Furthermore, although \textit{Hive} generally outperforms \textit{Illuminarty} on specific datasets such as Gemini-generated images ($\mu_{Hive}{=}0.892$ vs. $\mu_{Ill}{=}0.818$, $p < .001$, Cohen's d${=}0.281$), cross-platform agreements on individual images remains weak (Pearson $r{=}0.255$, ICC${=}0.231$, $p < .001$).



\begin{figure}[!htbp]
    \centering
    \includegraphics[width=0.5\textwidth]{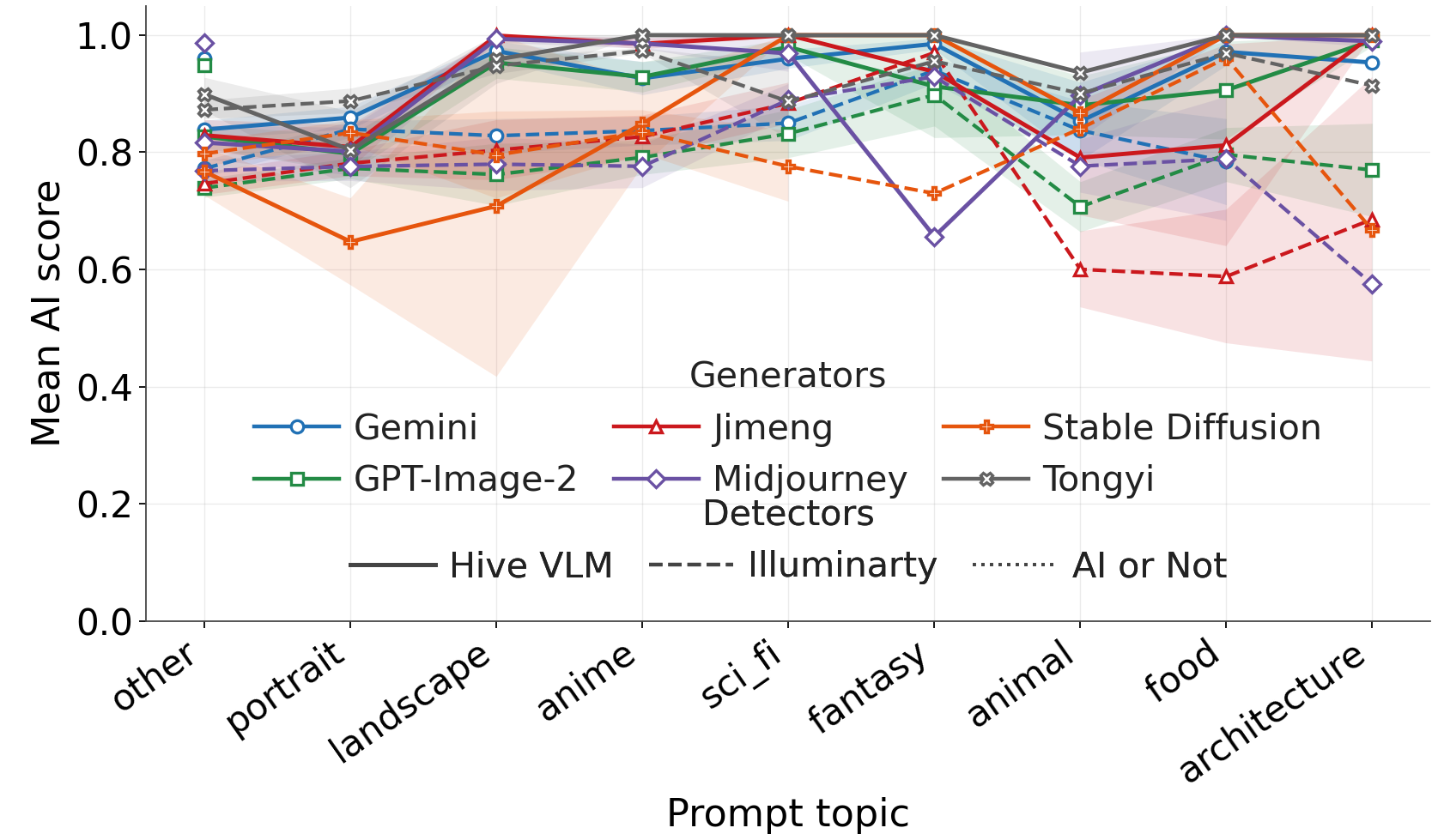}
    \caption{Detection accuracy across image topics and detector platforms.}
    \label{fig:topic}
\end{figure}

\begin{figure}[!htbp]
    \centering 
    \includegraphics[width=0.35\textwidth]{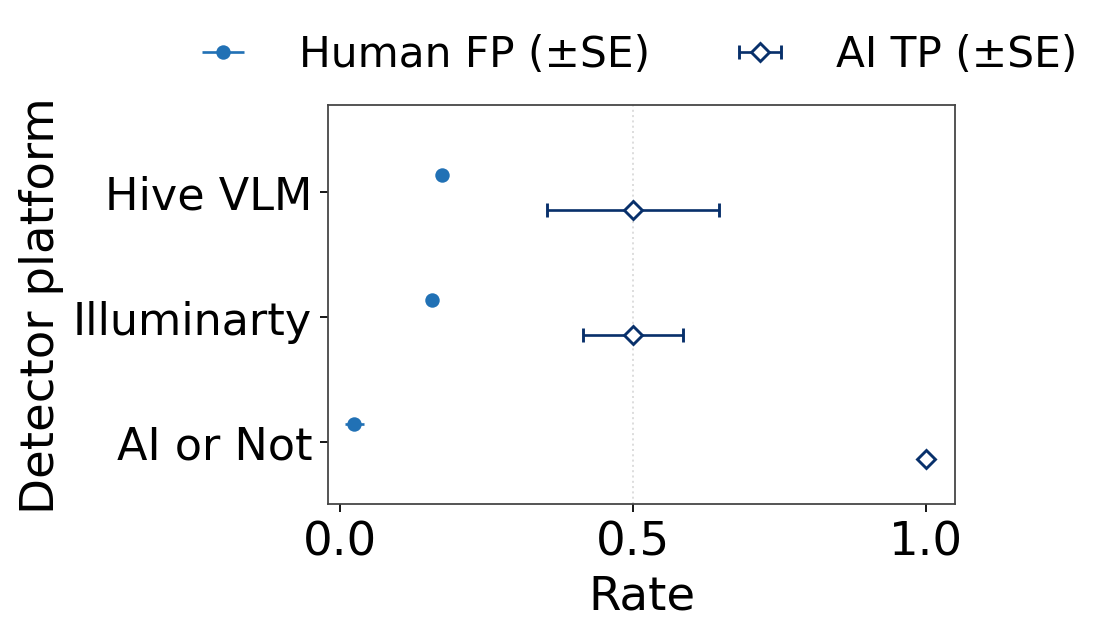}
    \caption{AI detection rates and false positive rates for human-authored images.}
    \label{fig:human_comparison}
\end{figure}

\textbf{Influence of image semantics and low-level features.} Detection efficacy varies across both semantic content of the prompt and the file's low-level properties, as in Figure~\ref{fig:topic}. For prompt topics, thematic content significantly impacts detection ($p{<}.001$, partial $\eta^2{=}0.023$). Notably, a significant performance gap exists between \textit{science fiction (sci\_fi)} and \textit{portrait} images ($p{<}.001$, Cohen's d${=}0.374$). Regarding low-level features, file size ($\rho{=}0.051$, $p{<}.001$) and color variance both have weak but significant correlations with detection scores. For false positive rates, when evaluating human-authored images such as from NASA and Wikimedia, \textit{Hive} incorrectly flags 17.5\% of images as AI-generated, significantly underperforming \textit{AI or Not}, which maintains a low 2.5\% false positive rate. This underscores the residual risk even on human-authored images, as in Figure~\ref{fig:human_comparison}. 



\begin{figure}[!htbp]
    \centering
    \includegraphics[width=0.47\textwidth]{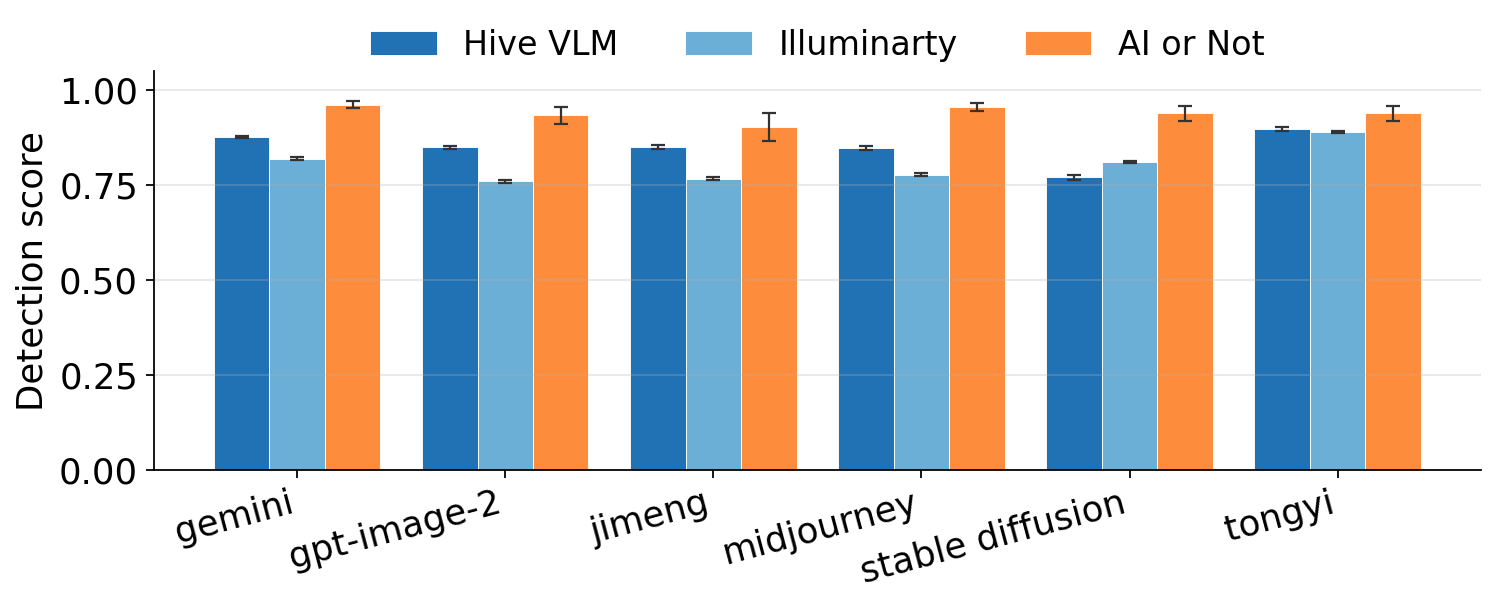}
    \caption{Detection scores across generation models and platforms.}
    \label{fig:generator_platforms}
\end{figure}

\textbf{Impact of AI generator.} 
The originality of AI generator significantly dictates detection success ($H{=}3,236$, $p < .001$, Figure~\ref{fig:generator_platforms}), with post-hoc corrections confirming significant differences in 13 of the 15 generator pairs ($\alpha$=0.05). Specifically, \textit{Tongyi} ($\mu=0.892$, SD$=0.215$), and \textit{Gemini} ($\mu=0.867$, median$=1.000$) yield higher detection scores, while \textit{Stable Diffusion} ranks the lowest ($\mu=0.787$).




\subsection{Metadata Detection}\label{sec:metadata_detection}

\textbf{Metadata detection without manipulations.} C2PA detection finds markers in 55.0\% of all corpus images (95\% CI [53.58\%, 56.42\%]). Manifest parsing yields a similar 54.94\% detection rate (95\% CI [53.52\%, 56.35\%]), while trained-algorithmic-media (TAM) flags identify 54.71\% (95\% CI [53.29\%, 56.12\%]). In contrast, SynthID markers remain absent (0.00\%; 95\% CI [0.00\%, 0.08\%]). We found rare disagreements between C2PA scanning and Manifest detection (McNemar exact: $p = .25$).

Among generated images, C2PA rates differ sharply by model ($\chi^2_5{=}3{,}276.94$, $p < .001$). Gemini achieves a 100.00\% detection rate (95\% CI [99.53\%, 100.00\%]), significantly outperforming GPT-Image-2 at 96.50\% (95\% CI [95.13\%, 97.50\%], $z = -5.41$, $p < .001$). Jimeng and Midjourney both cluster near 95\%, whereas Tongyi Wanxiang drops to 1.61\% (95\% CI [0.82\%, 3.14\%]) and Stable Diffusion shows 0.00\%. Pairwise contrasts confirm that Jimeng and Midjourney perform similarly ($z = -0.04$, $p = .971$), while Stable Diffusion and Tongyi significantly trail Gemini ($p < .001$).

\begin{figure}[!htbp]
    \centering
    \includegraphics[width=0.47\textwidth]{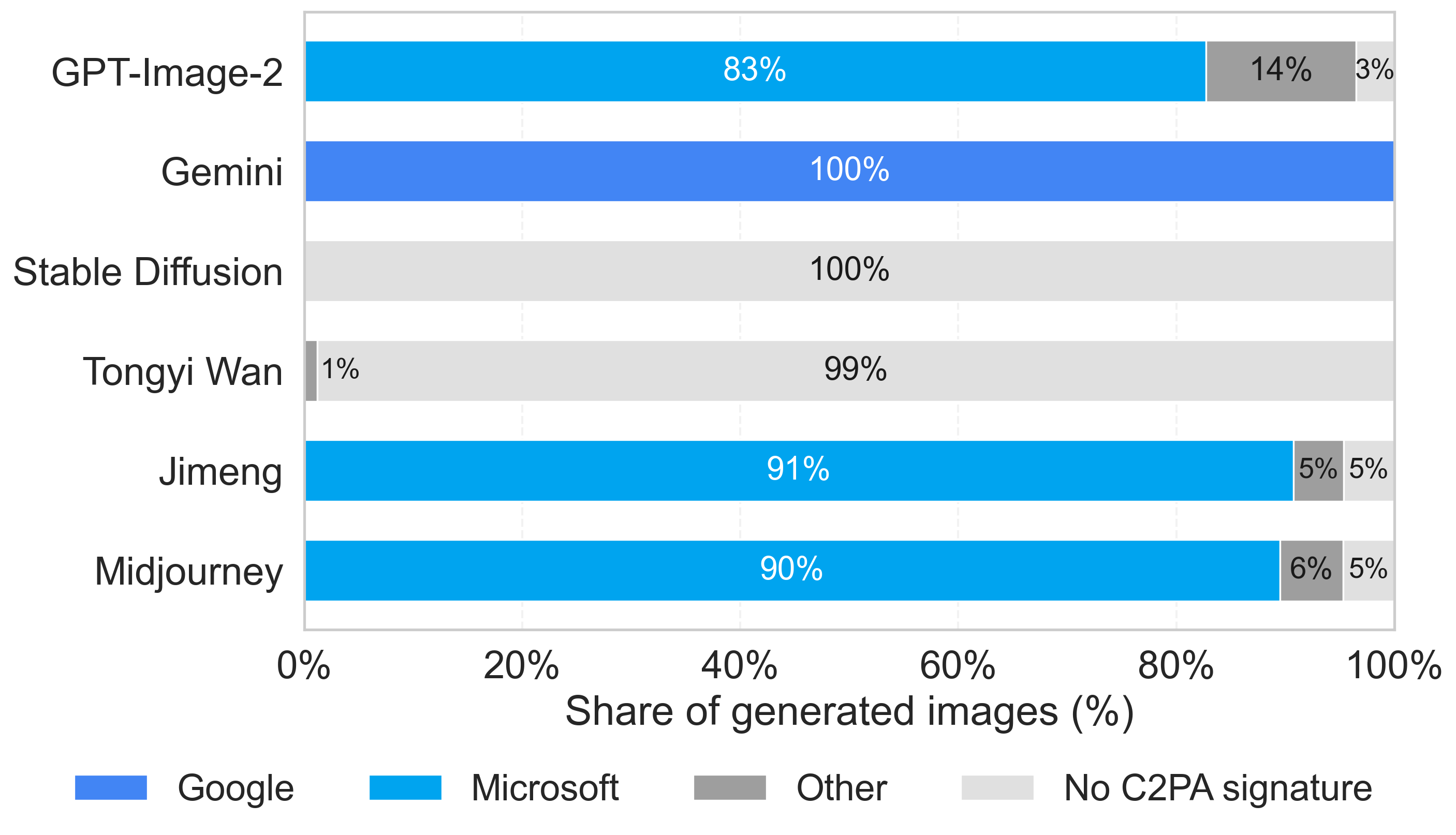}
    \caption{Stacked bar chart of C2PA signature issuer family by generator.}
    \label{fig:c2pa_issuer_stacked_by_generator}
\end{figure}

The internal signature structure also varies by generator ($\chi^2_{18} = 7{,}162.64$, $p < .001$, Figure~\ref{fig:c2pa_issuer_stacked_by_generator}). Microsoft commands a 43.5\% issuer family share, followed by Google (22.4\%), other issuers (5.0\%), and unsigned images (29.1\%). Regarding actions, \textit{created+watermarked} accounts for 30.3\%, \textit{created+edited} for 22.4\%, and \textit{created} for 17.9\%. Gemini outputs maintain a 100\% TAM flag rate. For GPT-Image-2 images with valid C2PA, TAM reaches 100\% and soft-binding reaches 49.3\%, while soft-binding and InvisMark each appear in 36.4\% of all generated cases. 

C2PA detection rates exhibit discrepancies across different generators. Gemini shows a 100.00\% detection rate (95\% CI [99.53\%, 100.00\%]), followed closely by GPT-Image-2 at 96.50\% (95\% CI [95.13\%, 97.50\%]), Jimeng at 95.36\% (95\% CI [93.02\%, 96.95\%]), and Midjourney at 95.31\% (95\% CI [92.94\%, 96.91\%]). Conversely, detection is nearly non-existent for Tongyi at 1.61\% and Stable Diffusion at 0.00\%. 

\begin{figure}[!htbp]
    \centering
    \includegraphics[width=0.47\textwidth]{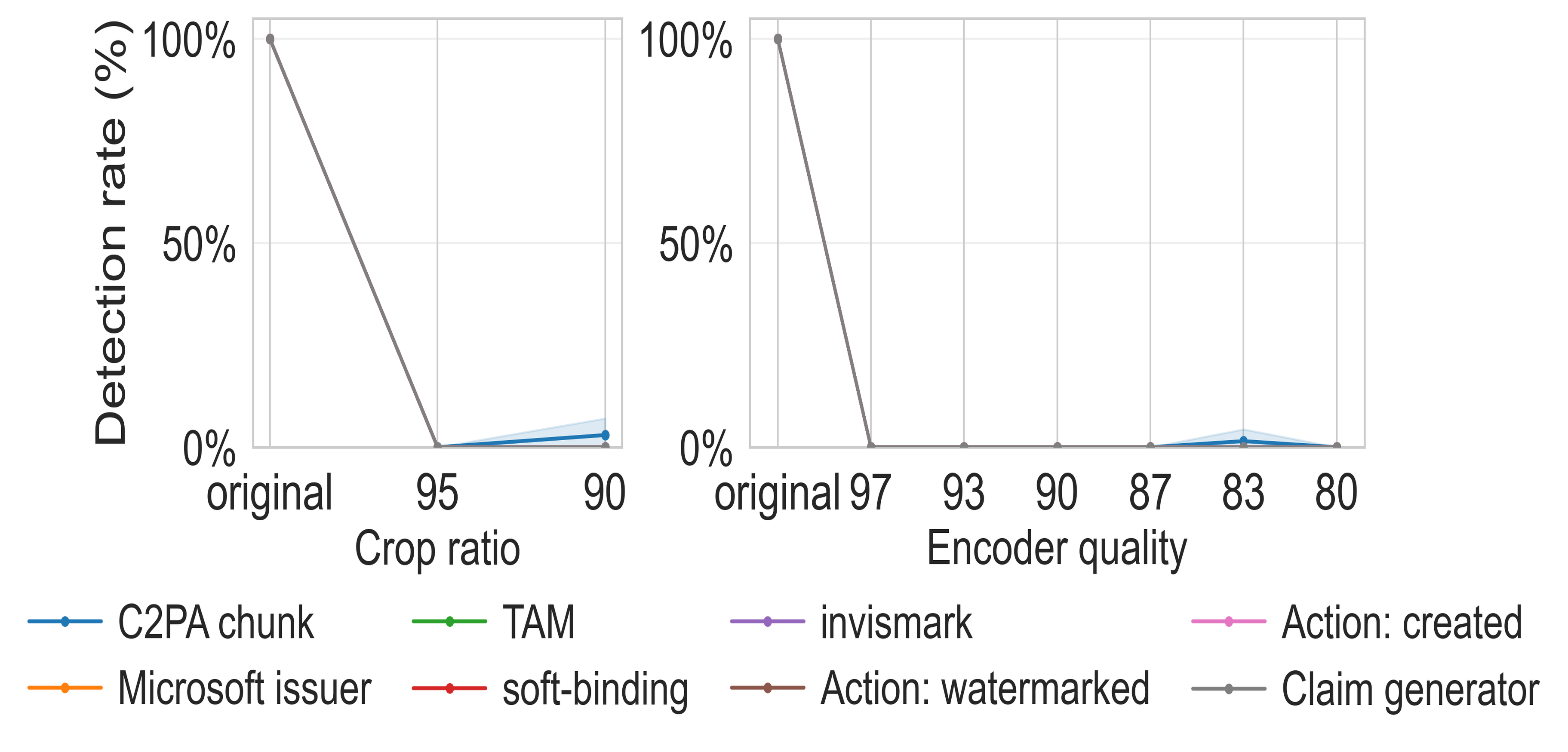}
    \caption{Retention rates across manifest attributes for each manipulation.}
    \label{fig:c2pa_field_survival}
\end{figure}

\textbf{Detection rates with manipulations.}
Image manipulations severely destroy metadata detections. Across all images, C2PA detection rates plummet from 71.0\% to just 0.47\% after manipulation (95\% CI [0.20\%, 1.09\%]; two-proportion $z = 40.81$, $p < .001$). Manifest data shows an identical drop to 0.37\% ($p < .001$). For GPT-Image-2 specifically, overall detection rates stay uniformly low across all manipulation types ($\chi^2_{15} = 17.48$, $p = .291$). Center-crop manipulations leave markers in only 1.00\% of images (95\% CI [0.27\%, 3.55\%]), while JPEG recompression retains 0.21\% (95\% CI [0.04\%, 1.20\%]) and quantization retains 0.75\% (95\% CI [0.20\%, 2.68\%]). Statistical tests show no significant differences between cropping and JPEG ($p = .165$), or cropping and quantization ($p = .772$). 
McNemar tests confirm that manipulations cause significant drop for \textit{c2pa\_present}, TAM ($p < .001$), soft-binding, and InvisMark ($p = .031$). As Figure~\ref{fig:c2pa_field_survival} shows, manipulations completely erase these structured attributes. Upon cropping, the \textit{action\_pattern} field shifts entirely to \textit{none}. Microsoft-issuer marks, TAM, soft-binding, and InvisMark declarations all drop to zero, except under the mild \textit{crop\_90} condition. Consequently, 99.6\% of manipulated GPT-Image-2 files lose their issuer data and carry a \textit{none} action label, stripping the rich corporate metadata present at generation. Besides, center cropping removes C2PA manifest chunks from the file, causing detection failure for manifest parsing.


\textbf{Robustness comparison with watermarks.} The fragility of metadata contrasts with the high detection rates of digital watermarks under the same manipulation settings. 

For embedded images, after applying \textit{Stable Signature}, the watermark achieves a 100.00\% \textit{detect\_ok} rate (95\% CI [99.89\%, 100.00\%]) under \textit{crop\_090} and \textit{crop\_095} conditions. JPEG recompression also preserves binary detection perfectly at 100.00\%, showing no statistical difference from cropping ($p = 1.0$). McNemar tests confirm zero discordant pairs between original and cropped images across all generators. However, spatial resizing severely degrades detection effectiveness, where a \textit{resize\_050} manipulation drops the detection rate to 0.00\% (95\% CI [0.00\%, 0.11\%], $p < .001$), lowering median scores from 1.0 to 0.792 ($p < .001$).

\begin{figure}[!htbp]
    \centering
    \includegraphics[width=0.47\textwidth]{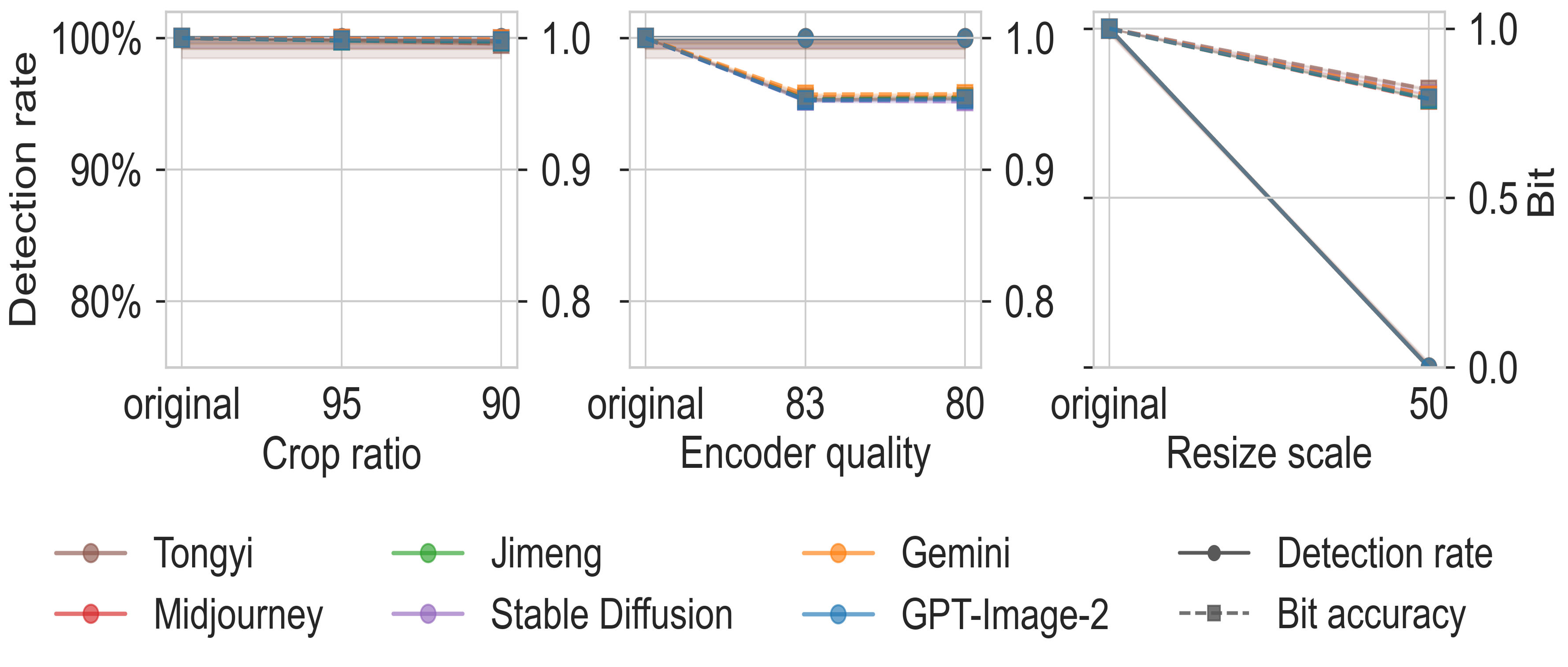}
    \caption{Stable Signature detection rate by generating models.}
    \label{fig:stable_signature_by_generator}
\end{figure}

Using Jonathan as an alternative detection tool further highlights this gap. On unmanipulated generated images, this tool results in a 100.00\% detection rate, while the detection rates of its C2PA and IPTC tags (\textit{iptc\_is\_ai\_generated}) are both at 94.41\% (95\% CI [90.02\%, 96.94\%]). Following cropping or JPEG compression attacks, the C2PA and IPTC metadata layers completely disappear to 0.00\% (95\% CI [0.00\%, 16.07\%]). Meanwhile, the pixel-embedded watermark signal maintains a perfect 100.00\% detection rate (Fisher's exact test, $p < .001$), suggesting that invisible watermarks are robust with different manipulations that even cause metadata to disappear during detection.



\begin{figure}[!htbp]
    \centering
    \includegraphics[width=0.47\textwidth]{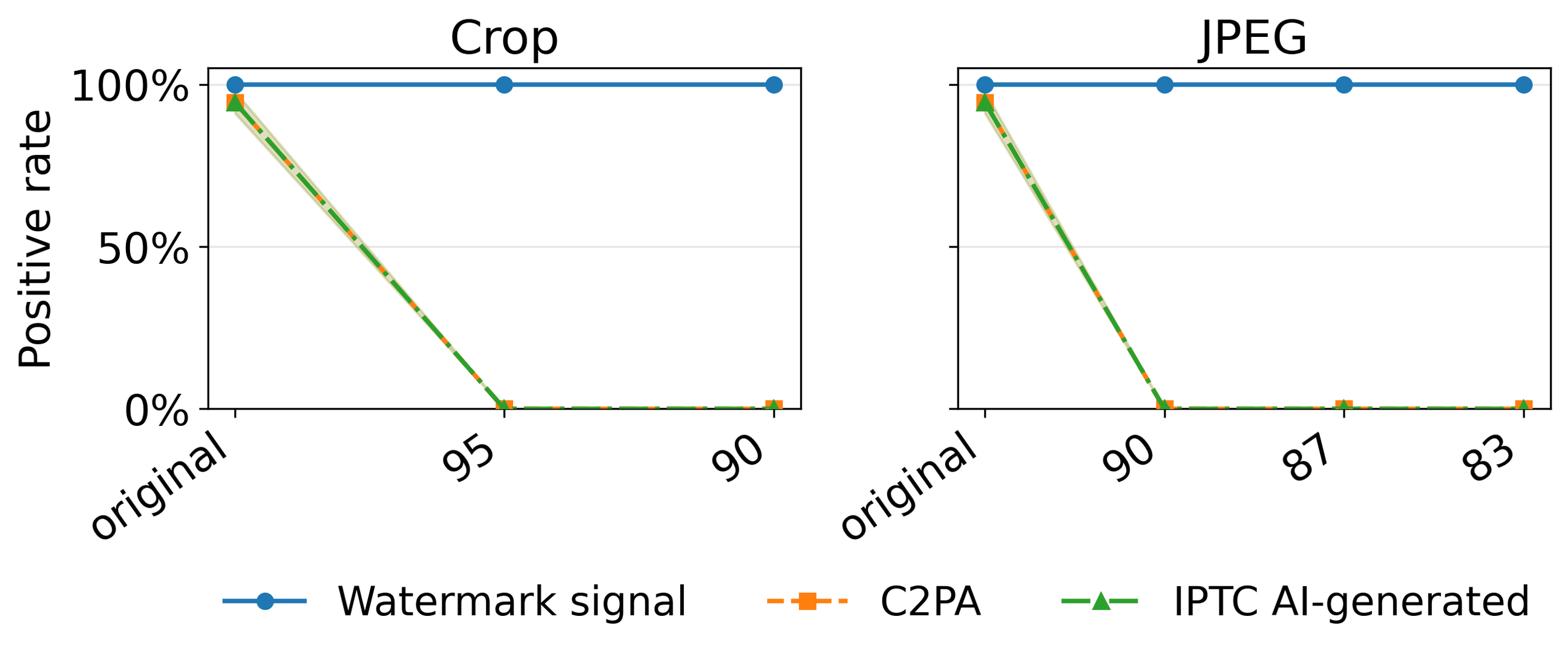}
    \caption{Jonathan Clark signal rates by modification.}
    \label{fig:jonathanclark_signals}
\end{figure}

\begin{framed}
\noindent \textbf{Takeaways:}

$\bullet$ 1. Commercial AI detectors resist most modifications but degrade under quantization. 

$\bullet$ 2. Cross-platform AI detectors' detection results are highly inconsistent.

$\bullet$ 3. Commercial AI detectors have high false-positive rates, up to 17.5\% on human-authentic images.

$\bullet$ 4. Metadata are highly vulnerable to manipulations, such as center-cropping or compression.

$\bullet$ 5. Invisible watermarking shows resilience against manipulations like cropping and compression.
\end{framed}
 
\section{Discussion}

\subsection{Discrepancies Between Creator Perceptions and Technical Realities}

By synthesizing qualitative interviews with quantitative analysis, we examine the nuances and misalignments between creators' perceptions regarding AI labels, and the actual performance of these traces in technical testing.

First, we find a mismatch between creators' strategies to evade platform detections, and the actual resilience of AI detectors. From interviews, creators wanted to evade downstream AI detectors. They perform edits such as center-cropping or screenshotting, and believe these actions can help them avoid platform detections, and thereby penalties like traffic suppression. However, our testing in RQ3 shows that commercial detectors resist most modifications, and their detection capabilities only drop significantly under severe quantization. 
Additionally, while most creators perform unified practices on these images across platforms, empirical analysis found different commercial tools give highly inconsistent results for the same images.

Second, we find that creators focus on visible elements when removing digital traces. Whereas in reality, the survivability of invisible digital traces under manipulations are inconsistent. From interviews, creators mostly focus on deleting visible marks and logos. However, empirical analysis found invisible watermarking show strong resilience against manipulations such as cropping and compression. Conversely, those invisible metadata are vulnerable to manipulations, where manipulations such as center-cropping or recompression cause detection rates to severely degrade.

Third, our tests found that current commercial detectors have false positives, misclassifying human-authored images as AI-generated. This may further exacerbate creators' anxieties regarding potential misjudgments and platform penalties, highlighting vulnerabilities within the current commercial AI detection landscape. 

\subsection{Uniqueness of Creators' Perspectives}

Unlike viewers who primarily evaluate content authenticity based on the presence of AI labels, creators place greater emphasis on autonomy, authenticity, and IP protection. Prior research from the viewer's perspective indicates a preference for conspicuous, platform-mandated AI labels~\cite{rae2024effects,jung2025ai}. In contrast, our findings suggest that compulsory and poorly calibrated label mechanisms can disincentivize generative tools' adoption and induce creators' manipulative strategies, ultimately undermining platform governance. 

Furthermore, while AI providers and researchers often focus on technical scalability and information load, creators prioritize their own agency and the potential S\&P implications of AI labels. Consequently, relying solely on technical dimensions is insufficient. Providers must balance transparency with user agency when enforcing AI labels.

Finally, traditional S\&P indicators, such as pre-AI era watermark techniques, primarily serve to protect system integrity. While they share the core functionality of IP protection with modern AI labels, they differ significantly in application. Specifically, AI labels are tasked with distinguishing human-authored images from AI-edited content, introducing distinct consequences if misused or poorly received by creators. Therefore, when interacting with AI-specific indicators, participants face novel trade-offs that necessitate more granular and sophisticated label frameworks.

\subsection{Implications}

\textbf{Evolving beyond binary indicators to address creator de-valuation fears.} We found that creators conflate binary AI labels with devaluation of their creative labor, driving them to remove watermarks to protect their professional credibility. 
This suggests that the problem is not disclosure itself, but that binary labels flatten diverse human-AI workflows into an ``AI-generated'' category, devaluing meaningful human contribution and reinforcing stigma around AI-assisted creation.
Therefore, future provenance systems should transition from binary AI-generated labels towards multi-tiered frameworks that quantify collaborative contributions. By measuring and displaying granular human editing traces, such as manual refinement and custom post-processing, platforms can eliminate ``all-or-nothing'' penalty and incentivize creators to voluntarily maintain AI labels.

\textbf{Deploying workflow-resilient implicit tracing mechanisms.} Platform operators, when employing S\&P tools, should prioritize multi-bit implicit watermarks over fragile manifest metadata. Our empirical test reveals that standard metadata structures like C2PA are completely wiped out under creator-reported manipulations such as center-cropping or JPEG recompression. To survive diverse real-world manipulations, AI labels need architectures that natively withstand pixel-level quantization and spatial editing transformations.

\textbf{Mitigating creator S\&P risks through identity-decoupled verification.} S\&P developers should decouple content accountability from personal identifier exposure by implementing privacy-preserving cryptographic protocols. We show that creators experience profound anxiety regarding de-anonymization, fearing that platform-embedded watermarks can link disjointed artworks back to their real-world identities, IP addresses, or contact information. Future standards should use anonymous credentials to confirm content provenance at the platform level, or in an anonymous manner without individual privacy leakage.

\textbf{Improving detection capabilities of commercial detectors.} Among different tested commercial detectors, their real-world efficacy are inconsistent. Therefore, the tech industry should establish unified standards to solve disparities across different commercial AI detectors. Further research needs to focus on building resilient cross-platform detection methods, preventing creators from suffering false-positive shadowbanning caused by uncalibrated errors.





\subsection{Limitations}

We acknowledged several limitations of this paper, and we mitigated these limitations as follows. \textbf{First, regarding demographics and methodology,} we primarily recruit creators from China and Germany, and relies on qualitative interviews, which are subject to self-reporting and recall bias. To mitigate these, we enforced rigorous screening criteria to ensure our participants have verified, diverse experience across AIGC platforms. Rather than claiming statistical generalization, we presented qualitative insights within their workflows.
\textbf{Second, regarding technical evaluation,} we focused on user-reported manipulation and commercial platforms, prioritizing ecological validity, omitting specialized watermark detection services or academic algorithms. Our tests are not meant to be comprehensive, and we advocate for future work to conduct a comprehensive benchmark.
\textbf{Third, regarding image operations,} our evaluation decoupled manipulation types to isolate their effects, while noting that future work could composite different manipulations to verify AI labels' survivability in complex workflows.



\section{Conclusion}

While AI labels are crucial for digital provenance and countering disinformation, their technical necessity conflicts with creators' reluctance to disclose AI assistance. Through semi-structured interviews (N=21) and robustness evaluations of images from six generative platforms against 16 manipulation workflows, we investigate how creators navigate this tension and how their subsequent actions alter AI labels. Our findings show that creators frequently misinterpret AI labels and fear de-anonymization via platform identifiers. Driven by concerns over algorithmic penalization and reputational loss, they actively remove these markers. Our measurements reveal that color quantization severely degrade label detection, while commercial detectors exhibit inconsistent performance, with true-positive rates varying from 48.8\% to 99.4\%. We therefore advocate for workflow-resilient AI labels that align technical security guarantees with creators' operational incentives.

\bibliographystyle{IEEEtran}
\bibliography{IEEEabrv,main}

\begin{thebibliography}{10}
\providecommand{\url}[1]{#1}
\csname url@samestyle\endcsname
\providecommand{\newblock}{\relax}
\providecommand{\bibinfo}[2]{#2}
\providecommand{\BIBentrySTDinterwordspacing}{\spaceskip=0pt\relax}
\providecommand{\BIBentryALTinterwordstretchfactor}{4}
\providecommand{\BIBentryALTinterwordspacing}{\spaceskip=\fontdimen2\font plus
\BIBentryALTinterwordstretchfactor\fontdimen3\font minus \fontdimen4\font\relax}
\providecommand{\BIBforeignlanguage}[2]{{%
\expandafter\ifx\csname l@#1\endcsname\relax
\typeout{** WARNING: IEEEtran.bst: No hyphenation pattern has been}%
\typeout{** loaded for the language `#1'. Using the pattern for}%
\typeout{** the default language instead.}%
\else
\language=\csname l@#1\endcsname
\fi
#2}}
\providecommand{\BIBdecl}{\relax}
\BIBdecl

\bibitem{abdelnabi2021adversarial}
S.~Abdelnabi and M.~Fritz, ``Adversarial watermarking transformer: Towards tracing text provenance with data hiding,'' in \emph{2021 IEEE symposium on security and privacy (SP)}.\hskip 1em plus 0.5em minus 0.4em\relax IEEE, 2021, pp. 121--140.

\bibitem{guo2025audio}
H.~Guo, J.~Guo, B.~Chen, Y.~Wang, X.~Chen, H.~Huang, Q.~Yan, and L.~Xiao, ``$\{$AUDIO$\}$$\{$WATERMARK$\}$: Dynamic and harmless watermark for black-box voice dataset copyright protection,'' in \emph{34th USENIX Security Symposium (USENIX Security 25)}, 2025, pp. 4601--4620.

\bibitem{qu2025provably}
W.~Qu, W.~Zheng, T.~Tao, D.~Yin, Y.~Jiang, Z.~Tian, W.~Zou, J.~Jia, and J.~Zhang, ``Provably robust multi-bit watermarking for $\{$AI-generated$\}$ text,'' in \emph{34th USENIX Security Symposium (USENIX Security 25)}, 2025, pp. 201--220.

\bibitem{rae2024effects}
I.~Rae, ``The effects of perceived ai use on content perceptions,'' in \emph{Proceedings of the 2024 CHI Conference on Human Factors in Computing Systems}, 2024, pp. 1--14.

\bibitem{jung2025ai}
Y.~Jung, P.~Hua, J.~Bao, and S.~S. Sundar, ``Ai-generated or ai-modified? user reactions to labeling ai use in social media posts,'' in \emph{Proceedings of the Extended Abstracts of the CHI Conference on Human Factors in Computing Systems}, 2025, pp. 1--7.

\bibitem{zhang2024confrontation}
S.~Zhang and S.~Li, ``"confrontation or acceptance": Understanding novice visual artists' perception towards ai-assisted art creation,'' \emph{arXiv preprint arXiv:2410.14925}, 2024.

\bibitem{messer2024co}
U.~Messer, ``Co-creating art with generative artificial intelligence: Implications for artworks and artists,'' \emph{Computers in human behavior: artificial humans}, vol.~2, no.~1, p. 100056, 2024.

\bibitem{kasra2018seeing}
M.~Kasra, C.~Shen, and J.~F. O'Brien, ``Seeing is believing: How people fail to identify fake images on the web,'' in \emph{Extended abstracts of the 2018 CHI conference on human factors in computing systems}, 2018, pp. 1--6.

\bibitem{wolf2026seeing}
E.~Wolf, Y.~Samaradivakara, O.~Gokhale, S.~Ahmed, Y.~Wang, P.~Pataranutaporn, and P.~Maes, ``Seeing is not believing: Realistic ai videos disrupt confidence in authentic videos and perceived reality,'' in \emph{Proceedings of the Extended Abstracts of the 2026 CHI Conference on Human Factors in Computing Systems}, 2026, pp. 1--10.

\bibitem{metaai2023stablesignature}
{Meta AI}, ``Stable signature: A new method for watermarking images created by open source generative ai,'' \url{https://ai.meta.com/blog/stable-signature-watermarking-generative-ai/}, 2023, [Accessed: 2026-06-11].

\bibitem{clark2025imagewatermark}
J.~Clark, ``Image watermark detection \& metadata extraction,'' \url{https://jonathanclark.com/posts/image-watermark-detection.html}, November 2025, [Accessed: 2026-06-11].

\bibitem{zhang2025agency}
\BIBentryALTinterwordspacing
S.~Zhang, H.~Wang, and X.~Yi, ``Exploring collaboration patterns and strategies in human-ai co-creation through the lens of agency: A scoping review of the top-tier hci literature,'' \emph{Proc. ACM Hum.-Comput. Interact.}, vol.~9, no.~7, Oct. 2025. [Online]. Available: \url{https://doi.org/10.1145/3757594}
\BIBentrySTDinterwordspacing

\bibitem{liu2022design}
V.~Liu and L.~B. Chilton, ``Design guidelines for prompt engineering text-to-image generative models,'' in \emph{Proceedings of the 2022 CHI conference on human factors in computing systems}, 2022, pp. 1--23.

\bibitem{chen2026generative}
C.~Chen, F.~Cheng, B.~Zhang, R.~Jin, C.~Dong, Z.~Sun, and Y.~Zhou, ``A generative ai-driven industrial design framework for human--genai co-creation,'' \emph{Symmetry}, vol.~18, no.~2, p. 352, 2026.

\bibitem{wan2024felt}
Q.~Wan, S.~Hu, Y.~Zhang, P.~Wang, B.~Wen, and Z.~Lu, ``"it felt like having a second mind": Investigating human-ai co-creativity in prewriting with large language models,'' \emph{Proceedings of the ACM on human-computer interaction}, vol.~8, no. CSCW1, pp. 1--26, 2024.

\bibitem{howison2011validity}
J.~Howison, A.~Wiggins, and K.~Crowston, ``Validity issues in the use of social network analysis with digital trace data,'' \emph{Journal of the Association for Information Systems}, vol.~12, no.~12, p.~2, 2011.

\bibitem{young2026media}
J.~Young, S.~Vaughan, A.~Jenks, H.~Malvar, C.~Paquin, P.~England, T.~Roca, J.~L. Ferres, F.~Poursabzi, N.~Coles \emph{et~al.}, ``Media integrity and authentication: Status, directions, and futures,'' \emph{arXiv preprint arXiv:2602.18681}, 2026.

\bibitem{nemecek2026authenticated}
A.~Nemecek, H.~He, G.~Cheng, and E.~Ayday, ``Authenticated contradictions from desynchronized provenance and watermarking,'' in \emph{Proceedings of the IEEE/CVF Conference on Computer Vision and Pattern Recognition}, 2026, pp. 10\,738--10\,748.

\bibitem{bistron2026deep}
M.~Bistro{\'n}, J.~M. {\.Z}urada, and Z.~Piotrowski, ``Deep learning for image watermarking: A comprehensive review and analysis of techniques, challenges, and applications,'' \emph{Sensors (Basel, Switzerland)}, vol.~26, no.~2, p. 444, 2026.

\bibitem{hwang2023brief}
J.~Hwang and S.~Oh, ``A brief survey of watermarks in generative ai,'' in \emph{2023 14th International Conference on Information and Communication Technology Convergence (ICTC)}.\hskip 1em plus 0.5em minus 0.4em\relax IEEE, 2023, pp. 1157--1160.

\bibitem{darwish2024blockchain}
S.~M. Darwish, M.~M. Abu-Deif, and S.~M. Elkaffas, ``Blockchain for video watermarking: An enhanced copyright protection approach for video forensics based on perceptual hash function,'' \emph{PloS one}, vol.~19, no.~10, p. e0308451, 2024.

\bibitem{meng2026advances}
J.~Meng and Z.~Lu, ``Advances in semantic-preserving text watermarking,'' \emph{Sensors}, vol.~26, no.~5, p. 1528, 2026.

\bibitem{bushey2025cryptographic}
J.~Bushey, N.~Rivard, and M.~Barbeau, ``Cryptographic provenance and ai-generated images,'' in \emph{2025 IEEE International Conference on Big Data (BigData)}.\hskip 1em plus 0.5em minus 0.4em\relax IEEE, 2025, pp. 5960--5967.

\bibitem{nin2013digital}
J.~Nin and S.~Ricciardi, ``Digital watermarking techniques and security issues in the information and communication society,'' in \emph{2013 27th International Conference on Advanced Information Networking and Applications Workshops}.\hskip 1em plus 0.5em minus 0.4em\relax IEEE, 2013, pp. 1553--1558.

\bibitem{zhao2021structural}
X.~Zhao, Y.~Yao, H.~Wu, and X.~Zhang, ``Structural watermarking to deep neural networks via network channel pruning,'' in \emph{2021 IEEE International Workshop on Information Forensics and Security (WIFS)}.\hskip 1em plus 0.5em minus 0.4em\relax IEEE, 2021, pp. 1--6.

\bibitem{lv2024mea}
P.~Lv, H.~Ma, K.~Chen, J.~Zhou, S.~Zhang, R.~Liang, S.~Zhu, P.~Li, and Y.~Zhang, ``Mea-defender: A robust watermark against model extraction attack,'' in \emph{2024 IEEE Symposium on Security and Privacy (SP)}.\hskip 1em plus 0.5em minus 0.4em\relax IEEE, 2024, pp. 2515--2533.

\bibitem{pegoraro2024deepeclipse}
A.~Pegoraro, C.~Segna, K.~Kumari, and A.-R. Sadeghi, ``$\{$DeepEclipse$\}$: How to break $\{$White-Box$\}$$\{$DNN-Watermarking$\}$ schemes,'' in \emph{33rd USENIX Security Symposium (USENIX Security 24)}, 2024, pp. 5287--5304.

\bibitem{zong2025audiomarknet}
W.~Zong, Y.-W. Chow, W.~Susilo, J.~Baek, and S.~Camtepe, ``$\{$AudioMarkNet$\}$: Audio watermarking for deepfake speech detection,'' in \emph{34th USENIX Security Symposium (USENIX Security 25)}, 2025, pp. 4663--4682.

\bibitem{bin2017study}
M.~A.~F. bin Jeffry and H.~K. Mammi, ``A study on image security in social media using digital watermarking with metadata,'' in \emph{2017 IEEE conference on application, information and network security (AINS)}.\hskip 1em plus 0.5em minus 0.4em\relax IEEE, 2017, pp. 118--123.

\bibitem{munier2013legal}
M.~Munier, V.~Lalanne, P.-Y. Ardoy, and M.~Ricarde, ``Legal issues about metadata data privacy vs information security,'' in \emph{International Workshop on Data Privacy Management}.\hskip 1em plus 0.5em minus 0.4em\relax Springer, 2013, pp. 162--177.

\bibitem{ulrich2022understanding}
H.~Ulrich, A.-K. Kock-Schoppenhauer, N.~Deppenwiese, R.~G{\"o}tt, J.~Kern, M.~Lablans, R.~W. Majeed, M.~R. St{\"o}hr, J.~Stausberg, J.~Varghese \emph{et~al.}, ``Understanding the nature of metadata: systematic review,'' \emph{Journal of medical Internet research}, vol.~24, no.~1, p. e25440, 2022.

\bibitem{burns2021making}
A.~Burns, T.~On, C.~Lee, R.~Shapiro, C.~Xiong, and N.~Mahyar, ``Making the invisible visible: Risks and benefits of disclosing metadata in visualization,'' in \emph{2021 IEEE Workshop on Visualization for Social Good (VIS4Good)}.\hskip 1em plus 0.5em minus 0.4em\relax IEEE, 2021, pp. 11--15.

\bibitem{zhang2025hard}
Y.~Zhang, B.~Shi, S.~Qi, X.~Xiao, P.~Wang, and W.~Wen, ``Hard exif: Protecting image authorship through metadata, hardware, and content,'' \emph{IEEE Transactions on Image Processing}, 2025.

\bibitem{drakonakis2019please}
K.~Drakonakis, P.~Ilia, S.~Ioannidis, and J.~Polakis, ``Please forget where i was last summer: The privacy risks of public location (meta) data,'' \emph{arXiv preprint arXiv:1901.00897}, 2019.

\bibitem{kumar2016location}
P.~R. Kumar, C.~Srikanth, and K.~Sailaja, ``Location identification of the individual based on image metadata,'' \emph{Procedia Computer Science}, vol.~85, pp. 451--454, 2016.

\bibitem{baron2020you}
B.~Baron and M.~Musolesi, ``Where you go matters: a study on the privacy implications of continuous location tracking,'' \emph{Proceedings of the ACM on Interactive, Mobile, Wearable and Ubiquitous Technologies}, vol.~4, no.~4, pp. 1--32, 2020.

\bibitem{mayer2016evaluating}
J.~Mayer, P.~Mutchler, and J.~C. Mitchell, ``Evaluating the privacy properties of telephone metadata,'' \emph{Proceedings of the National Academy of Sciences}, vol. 113, no.~20, pp. 5536--5541, 2016.

\bibitem{perez2018you}
B.~Perez, M.~Musolesi, and G.~Stringhini, ``You are your metadata: Identification and obfuscation of social media users using metadata information,'' in \emph{Proceedings of the International AAAI Conference on Web and Social Media}, vol.~12, no.~1, 2018.

\bibitem{zhan2024will}
D.~Zhan and R.~Hai, ``Will sharing metadata leak privacy?'' in \emph{2024 IEEE 40th International Conference on Data Engineering Workshops (ICDEW)}.\hskip 1em plus 0.5em minus 0.4em\relax IEEE, 2024, pp. 317--323.

\bibitem{liang2026watermarking}
Y.~Liang, J.~Xiao, W.~Gan, and P.~S. Yu, ``Watermarking techniques for large language models: A survey,'' \emph{Artificial Intelligence Review}, 2026.

\bibitem{zhao2024invisible}
X.~Zhao, K.~Zhang, Z.~Su, S.~Vasan, I.~Grishchenko, C.~Kruegel, G.~Vigna, Y.-X. Wang, and L.~Li, ``Invisible image watermarks are provably removable using generative ai,'' \emph{Advances in neural information processing systems}, vol.~37, pp. 8643--8672, 2024.

\bibitem{c2pa_advancing}
{Coalition for Content Provenance and Authenticity (C2PA)}, ``Advancing digital content transparency and authenticity,'' \url{https://c2pa.org/}, 2026, [Accessed: 2026-6-12].

\bibitem{steinebach2023analysis}
M.~Steinebach, ``An analysis of photodna,'' in \emph{Proceedings of the 18th International Conference on Availability, Reliability and Security}, 2023, pp. 1--8.

\bibitem{mohit2026provenance}
A.~Mohit, B.~Aggarwal, and C.~Gondhalekar, ``Provenance verification of ai-generated images via a perceptual hash registry anchored on blockchain,'' \emph{arXiv preprint arXiv:2602.02412}, 2026.

\bibitem{dathathri2024scalable}
S.~Dathathri, A.~See, S.~Ghaisas, P.-S. Huang, R.~McAdam, J.~Welbl, V.~Bachani, A.~Kaskasoli, R.~Stanforth, T.~Matejovicova \emph{et~al.}, ``Scalable watermarking for identifying large language model outputs,'' \emph{Nature}, vol. 634, no. 8035, pp. 818--823, 2024.

\bibitem{gowal2025synthid}
S.~Gowal, R.~Bunel, F.~Stimberg, D.~Stutz, G.~Ortiz-Jimenez, C.~Kouridi, M.~Vecerik, J.~Hayes, S.-A. Rebuffi, P.~Bernard \emph{et~al.}, ``Synthid-image: Image watermarking at internet scale,'' \emph{arXiv preprint arXiv:2510.09263}, 2025.

\bibitem{an2024waves}
B.~An, M.~Ding, T.~Rabbani, A.~Agrawal, Y.~Xu, C.~Deng, S.~Zhu, A.~Mohamed, Y.~Wen, T.~Goldstein \emph{et~al.}, ``Waves: benchmarking the robustness of image watermarks,'' in \emph{Proceedings of the 41st International Conference on Machine Learning}, 2024, pp. 1456--1492.

\bibitem{jovanovic2026watermarking}
N.~Jovanovi{\'c}, I.~Labiad, T.~Sou{\v{c}}ek, M.~Vechev, and P.~Fernandez, ``Watermarking autoregressive image generation,'' \emph{Advances in Neural Information Processing Systems}, vol.~38, pp. 71\,801--71\,848, 2026.

\bibitem{kassis2025unmarker}
A.~Kassis and U.~Hengartner, ``Unmarker: A universal attack on defensive image watermarking,'' in \emph{2025 IEEE Symposium on Security and Privacy (SP)}.\hskip 1em plus 0.5em minus 0.4em\relax IEEE, 2025, pp. 2602--2620.

\bibitem{shamshad2025first}
F.~Shamshad, T.~Bakr, Y.~S. Shaaban, N.~H. Hussein, K.~Nandakumar, and N.~Lukas, ``First-place solution to neurips 2024 invisible watermark removal challenge,'' in \emph{The 1st Workshop on GenAI Watermarking}, 2025.

\bibitem{zhao2023provable}
X.~Zhao, P.~V. Ananth, L.~Li, and Y.-X. Wang, ``Provable robust watermarking for ai-generated text,'' in \emph{The Twelfth International Conference on Learning Representations}, 2023.

\bibitem{moruzzi2025content}
C.~Moruzzi, E.~Tallyn, F.~Liddell, B.~Dixon, J.~Collomosse, and C.~Elsden, ``Content authenticities: A discussion on the values of provenance data for creatives and their audiences,'' in \emph{Proceedings of the 2025 Conference on Creativity and Cognition}, 2025, pp. 128--140.

\bibitem{dhawka2025data}
P.~Dhawka, N.~Lutz, and K.~Starbird, ``Data visualizations as propaganda: Tracing lineages, provenance, and political framings in online anti-immigrant discourse,'' \emph{Proceedings of the ACM on Human-Computer Interaction}, vol.~9, no.~7, pp. 1--47, 2025.

\bibitem{sunshine2009crying}
J.~Sunshine, S.~Egelman, H.~Almuhimedi, N.~Atri, and L.~F. Cranor, ``Crying wolf: An empirical study of ssl warning effectiveness.'' in \emph{USENIX security symposium}.\hskip 1em plus 0.5em minus 0.4em\relax Montreal, Canada, 2009, pp. 399--416.

\bibitem{krombholz2019if}
K.~Krombholz, K.~Busse, K.~Pfeffer, M.~Smith, and E.~Von~Zezschwitz, ``"if https were secure, i wouldn't need 2fa"-end user and administrator mental models of https,'' in \emph{2019 IEEE Symposium on security and privacy (SP)}.\hskip 1em plus 0.5em minus 0.4em\relax IEEE, 2019, pp. 246--263.

\bibitem{abu2018exploring}
R.~Abu-Salma, E.~M. Redmiles, B.~Ur, and M.~Wei, ``Exploring user mental models of $\{$End-to-End$\}$ encrypted communication tools,'' in \emph{8th USENIX Workshop on Free and Open Communications on the Internet (FOCI 18)}, 2018.

\bibitem{unger2015sok}
N.~Unger, S.~Dechand, J.~Bonneau, S.~Fahl, H.~Perl, I.~Goldberg, and M.~Smith, ``Sok: secure messaging,'' in \emph{2015 IEEE Symposium on Security and Privacy}.\hskip 1em plus 0.5em minus 0.4em\relax IEEE, 2015, pp. 232--249.

\bibitem{abu2020designing}
R.~Abu-Salma, ``Designing user-centered privacy-enhancing technologies,'' Ph.D. dissertation, UCL (University College London), 2020.

\bibitem{shahriari2026systematic}
R.~Shahriari and E.~D. Ragan, ``A systematic survey of empirical user studies of unintentional information disclosure in everyday digital interaction,'' \emph{International Journal of Human--Computer Interaction}, pp. 1--29, 2026.

\bibitem{tayeb2018toward}
S.~Tayeb, A.~Week, J.~Yee, M.~Carrera, K.~Edwards, V.~Murray-Garcia, M.~Marchello, J.~Zhan, and M.~Pirouz, ``Toward metadata removal to preserve privacy of social media users,'' in \emph{2018 IEEE 8th Annual Computing and Communication Workshop and Conference (CCWC)}.\hskip 1em plus 0.5em minus 0.4em\relax IEEE, 2018, pp. 287--293.

\bibitem{henne2014awareness}
B.~Henne, M.~Koch, and M.~Smith, ``On the awareness, control and privacy of shared photo metadata,'' in \emph{International Conference on Financial Cryptography and Data Security}.\hskip 1em plus 0.5em minus 0.4em\relax Springer, 2014, pp. 77--88.

\bibitem{jorgensen2023designing}
M.~N. J{\o}rgensen and T.~Jenkins, ``Designing anekdota: Investigating personal metadata for legacy,'' in \emph{Proceedings of the 2023 CHI Conference on Human Factors in Computing Systems}, 2023, pp. 1--14.

\bibitem{coleti2020tr}
T.~A. Coleti, P.~L.~P. Corr{\^e}a, L.~V.~L. Filgueiras, and M.~Morandini, ``Tr-model. a metadata profile application for personal data transparency,'' \emph{IEEE Access}, vol.~8, pp. 75\,184--75\,209, 2020.

\bibitem{henne2013snapme}
B.~Henne, C.~Szongott, and M.~Smith, ``Snapme if you can: Privacy threats of other peoples' geo-tagged media and what we can do about it,'' in \emph{Proceedings of the sixth ACM conference on Security and privacy in wireless and mobile networks}, 2013, pp. 95--106.

\bibitem{chen2024scalable}
Y.~Chen, D.~Heath, R.~Chatterjee, and E.~Fernandes, ``Scalable metadata-hiding for privacy-preserving iot systems,'' \emph{Proceedings on Privacy Enhancing Technologies}, 2024.

\bibitem{beard2018digital}
I.~Beard, ``Digital photos, embedded metadata and personal privacy,'' \emph{The Complete Guide to Personal Digital Archiving}, pp. 201--212, 2018.

\bibitem{furini2015location}
M.~Furini and V.~Tamanini, ``Location privacy and public metadata in social media platforms: attitudes, behaviors and opinions,'' \emph{Multimedia Tools and Applications}, vol.~74, no.~21, pp. 9795--9825, 2015.

\bibitem{xiao2025authorship}
Y.~Xiao, ``How authorship labels shape perceptions, attribution judgments, and willingness to pay for ai-human co-created music,'' in \emph{2025 IEEE International Symposium on Technology and Society (ISTAS)}.\hskip 1em plus 0.5em minus 0.4em\relax IEEE, 2025, pp. 1--6.

\bibitem{burrus2024unmasking}
O.~Burrus, A.~Curtis, and L.~Herman, ``Unmasking ai: Informing authenticity decisions by labeling ai-generated content,'' \emph{Interactions}, vol.~31, no.~4, pp. 38--42, 2024.

\bibitem{holtervennhoff2026s}
S.~H{\"o}ltervennhoff, J.~Ricker, M.~M. Raphael, C.~Schwedes, R.~Weil, A.~Fischer, T.~Holz, L.~Sch{\"o}nherr, and S.~Fahl, ``" that's another doom i haven't thought about": A user study on ai labels as a safeguard against image-based misinformation,'' in \emph{Proceedings of the 2026 CHI Conference on Human Factors in Computing Systems}, 2026, pp. 1--32.

\bibitem{gamageLabeling}
D.~Gamage, D.~Sewwandi, M.~Zhang, and A.~K. Bandara, ``Labeling synthetic content: User perceptions of label designs for ai-generated content on social media,'' in \emph{Proceedings of the 2025 CHI conference on human factors in computing systems}, 2025, pp. 1--29.

\bibitem{PanickerFraudulence}
A.~Panicker, N.~Nurain, Z.~Ibrahim, C.-H. Wang, S.~W. Ha, Y.~Wu, K.~Connelly, K.~A. Siek, and C.-F. Chung, ``Understanding fraudulence in online qualitative studies: from the researcher's perspective,'' in \emph{Proceedings of the 2024 CHI conference on human factors in computing systems}, 2024, pp. 1--17.

\bibitem{mcdonald2019reliability}
N.~McDonald, S.~Schoenebeck, and A.~Forte, ``Reliability and inter-rater reliability in qualitative research: Norms and guidelines for cscw and hci practice,'' \emph{Proceedings of the ACM on human-computer interaction}, no. CSCW, pp. 1--23, 2019.

\bibitem{deng2026prompt}
Z.~Deng, H.~Li, W.~Ma, R.~Sun, D.~Wang, M.~Xue, H.~Hu, S.~Wen, and Y.~Xiang, ``The prompt stealing fallacy: Rethinking metrics, attacks, and defenses,'' 2026.

\bibitem{shen2024prompt}
X.~Shen, Y.~Qu, M.~Backes, and Y.~Zhang, ``Prompt stealing attacks against $\{$Text-to-Image$\}$ generation models,'' in \emph{33rd USENIX Security Symposium (USENIX Security 24)}, 2024, pp. 5823--5840.

\bibitem{lukas2023ptw}
N.~Lukas and F.~Kerschbaum, ``$\{$PTW$\}$: Pivotal tuning watermarking for $\{$Pre-Trained$\}$ image generators,'' in \emph{32nd USENIX Security Symposium (USENIX Security 23)}, 2023, pp. 2241--2258.

\bibitem{wallace1991jpeg}
G.~K. Wallace, ``The jpeg still picture compression standard,'' \emph{Communications of the ACM}, vol.~34, no.~4, pp. 30--44, 1991.

\bibitem{liu2025screenguard}
G.~Liu, X.~Liang, X.~Hu, Y.~Si, X.~Zhang, and Z.~Qian, ``Screenguard: A screen-targeted watermarking scheme against arbitrary screenshot,'' \emph{IEEE Transactions on Multimedia}, 2025.

\bibitem{rao2025dynmark}
C.~Rao, G.~Liu, S.~Li, X.~Zhang, and Z.~Qian, ``Dynmark: A robust watermarking solution for dynamic screen content with small-size screenshot support,'' in \emph{Proceedings of the 33rd ACM International Conference on Multimedia}, 2025, pp. 7463--7471.

\bibitem{xiao2024client}
X.~Xiao, Y.~Zhang, Z.~Hua, Z.~Xia, and J.~Weng, ``Client-side embedding of screen-shooting resilient image watermarking,'' \emph{IEEE Transactions on Information Forensics and Security}, vol.~19, pp. 5357--5372, 2024.

\end{thebibliography}
%



\appendices

\section{Additional Details for the Detectors}

\textbf{C2PA scan.} We apply offline byte-level scanner that flags C2PA-related markers. Each image yields \textit{c2pa\_detected}, \textit{synthid\_detected} and \textit{any\_openai\_signal}. 

\textbf{C2PA manifest.} For every image we parse embedded payloads, and record claim-level fields (\textit{c2pa\_present}, \textit{claim\_generator}, \textit{software\_agent}), action metadata (\textit{action\_types}, \textit{action\_pattern}), \textit{digital\_source\_type}, \textit{has\_trained\_algorithmic\_media} (TAM), \textit{signature\_issuer\_family}, \textit{has\_soft\_binding}, \textit{has\_invismark}, and manifest size. 

\textbf{Stable signature.} For images, we re-embed Meta Stable Signature's hidden watermark after using models to generate the images. We detect the watermarks before and after manipulations. Detection reports \textit{detect\_ok} and per-bit \textit{bit\_accuracy}. 

\textbf{Jonathan Clark's Detector API.} We also chose the platform provided by Jonathan Clark\footnote{https://jonathanclark.com/}, a commercial detector, and used their API for detecting metadata. Detected signals include \textit{watermark\_signal\_detected}, \textit{c2pa\_detected}, and \textit{iptc\_is\_ai\_generated}. 

\textbf{Statistical analysis.} Proportions are reported as k/N with 95\% Wilson intervals. Generator-level heterogeneity is assessed using Pearson's $\chi^2$ tests, with two-proportion $z$-tests applied for pairwise generator contrasts. For paired comparisons between original and manipulated versions, we employ exact McNemar's tests. Finally, differences in continuous variables (e.g., bit accuracy, manifest size) across independent groups are evaluated using two-sided Mann-Whitney $U$ tests.

\section{Parameter Details of the Evaluation}\label{app:para_details}

We documented the evaluation details as follows.

(1) JPEG compression: following prior practices~\cite{lukas2023ptw}, we perform JPEG compression~\cite{wallace1991jpeg} on the base image with a quality $q$. A higher quality preserves the visual quality of the image better, and we experiment with $q \in [80, 100]$.

(2) Resaving to other formats: We tried resaving PNG image to JPG and SVG formats, as all platforms we tried save the image as PNG by default. When saving these images to JPG and SVG formats, we kept the quality by default, corresponding to a quality $q=100$ for JPG. This also avoids confounding with the JPEG compression experiment.
 
(3) Cropping: For this class, following prior practices~\cite{lukas2023ptw}, the base image is center-cropped with a given cropping ratio $\rho \in (0, 1]$, and then resizes the cropping back to the base image's original size. We experiment with cropping ratios $\rho \in [0.9, 1]$.

(4) Reposting and downloading: We reposted the pictures to Tiktok, Instagram, Facebook, X, Weixin Moment, Weixin Group, Xiaohongshu, Weibo, which represent popular social platforms. The picture is first posted and then downloaded, which tests whether platforms would add manipulations to the picture.

(5) Screenshotting: We tested screenshotting the picture using standard settings on laptops, mimicking normal workflows of AIGC creators. We screenshotted the images with the original resolution~\cite{liu2025screenguard,rao2025dynmark,xiao2024client}.

\section{Post-hoc Contrasts Versus Original Images}

Table~\ref{tab:rq3_manipulation_posthoc} shows the post-hoc contrasts versus original images. 

\begin{table}[!htbp]
\centering
\caption{Post-hoc contrasts versus original images.}
\label{tab:rq3_manipulation_posthoc}
\small
\begin{tabular}{@{}lrr@{}}
\toprule
Attack & $\Delta\mu$ & FDR $p$ \\
\midrule
\textit{Crop }$r{=}100$ & $+0.002$ & .837 \\
\textit{Crop }$r{=}95$  & $+0.001$ & .988 \\
\textit{Crop }$r{=}90$  & $-0.001$ & .988 \\
\textit{Screenshot (same)} & $+0.002$ & .988 \\
\textit{JPEG }$q{=}80$  & $-0.021$ & .282 \\
\textit{JPEG }$q{=}83$  & $-0.020$ & .282 \\
\textit{JPEG }$q{=}87$  & $-0.015$ & .587 \\
\textit{JPEG }$q{=}90$  & $-0.014$ & .612 \\
\textit{JPEG }$q{=}93$  & $-0.012$ & .702 \\
\textit{JPEG }$q{=}97$  & $-0.009$ & .708 \\
\textit{JPEG }$q{=}100$ & $-0.013$ & .612 \\
\textit{Quantize }$q{=}1$ & $+0.003$ & .837 \\
\textit{Quantize }$q{=}2$ & $+0.000$ & .988 \\
\textit{Quantize }$q{=}5$ & $+0.014$ & .282 \\
\textbf{Quantize }$q{=}10$ & \textbf{$+0.041$} & \textbf{$<$.001} \\
\bottomrule
\end{tabular}
\end{table}

\section{Demographic}
Table~\ref{tab:demographic} showed the demographic for this paper. 

\begin{table*}[!htbp] 
  \centering 
  \small
  
  \caption{Participant demographics, including topics of creation, tools used, creation frequency, and content modality.}
  \label{tab:demographic}
  
  \begin{tabularx}{\textwidth}{p{0.25cm} p{0.5cm} p{0.8cm} p{1cm} p{4cm} p{4cm} p{1.5cm} X}
    \toprule
    ID & Gender & Age & Education & Topic & Tools & Frequency & Modality \\
    \midrule
    P1  & F & 18--25 & Master & Film and television content & KLING, Midjourney, Stable Diffusion, JiMeng, ChatGPT & At least 3 times weekly & Images, Videos \\
    P2  & M & 26--35 & Master & Animation and visual exploration & KLING, JiMeng, Sora, Veo3, Vidu & Daily & Images, Videos \\
    P3  & F & 26--35 & Master & Creative production and early AI experiments & Midjourney & At least 3 times weekly & Text, Images \\
    P4  & F & 18--25 & Bachelor & MV, chibi illustration, and comic-style art & Doubao, JiMeng & At least 3 times weekly & Images, Videos \\
    P5  & M & 26--35 & Bachelor & Traffic-driven commercial content creation & DeepSeek, CapCut, Doubao & Daily & Images, Videos \\
    P6  & F & 26--35 & Master & Multi-format content production and client collaboration & Midjourney, DeepSeek & At least 3 times weekly & Images \\
    P7  & F & 18--25 & Bachelor & Entertainment-oriented and short-form videos & KLING, Midjourney, Vidu, Veo3, Jixiang AI & At least 3 times weekly & Videos \\
    P8  & F & 26--35 & Master & Music tools, competitions, AIGC animation, and article writing & Midjourney, Xiliu, JiMeng, DeepSeek, ChatGPT, Kimi, Coze, Suno & Daily & Videos, Audio \\
    P9  & F & 18--25 & Bachelor & Celebrity-related content, school assignments, and personal photo editing & JiMeng, Xingye, Doubao & At least 3 times weekly & Images, Videos, Audio \\
    P10 & F & 18--25 & Bachelor & AIGC-related competitions & KLING, Midjourney, Runway, Vidu & At least 3 times weekly & Images, Videos \\
    P11 & F & 18--25 & Bachelor & Food-related and ideological content & Doubao, JiMeng & At least 3 times weekly & Images, Videos \\
    P12 & M & 18--25 & Master & AI video creation for competitions and educational animation & Gemini, Claude, ChatGPT, Veo3, Vidu2, Lovart, Manus & Daily & Text, Code, Images, Videos, Audio \\
    P13 & F & 26--35 & Master & Psychological curriculum content and media production & KLING, Midjourney & Daily & Images, Videos \\
    P14 & F & 26--35 & Bachelor & Clothing and accessories & ChatGPT, KLING, JiMeng, Gemini & At least 3 times weekly & Images \\
    P15 & M & 18--25 & High school & Drawing and illustration & Sora, ChatGPT, Gemini, Perplexity & Daily & Images, Videos, Code \\
    P16 & F & 18--25 & Master & Animations, poetry, ringtones, and city-themed content & KLING, Midjourney & Daily & Images, Videos, Audio \\
    P17 & F & 18--25 & Bachelor & Animation, documentary, and war-related content & Midjourney, Sora, JiMeng, Doubao & At least 3 times weekly & Images \\
    P18 & F & 18--25 & Bachelor & Fan cards and character-focused content & CapCut & At least 3 times weekly & Images \\
    P19 & F & 18--25 & Master & Furniture design & Midjourney, ChatGPT, JiMeng, Gemini & Daily & Images, Videos \\
    P20 & M & 26--35 & Master & Promotional images, book covers, and product visuals & ChatGPT, Doubao, TikTok, Xingtu, Meitu & Daily & Text, Images, Audio \\
    P21 & F & 18--25 & Bachelor & Promotional images and fan fiction & Midjourney, Stable Diffusion, Sora & Daily & Text, Images \\
    \bottomrule
  \end{tabularx}
\end{table*}

\end{document}